\documentclass[a4paper,11pt,envcountsame]{llncs}%
\usepackage[utf8]{inputenc}%
\usepackage[IL2]{fontenc}%
\usepackage{a4wide}%
\usepackage{amssymb}%
\usepackage{amsmath}%
\usepackage{pstricks}%
\usepackage{pst-node}%
\usepackage{multido}%
\input{macros}
\begin{document}%

\title{Braess's Paradox for Flows Over Time} 
\author{%
  Martin Macko\inst{1}%
  \thanks{Supported in part by %
    the Comenius University grant No.~UK/358/2008, %
    the National Scholarship Programme of the Slovak Republic, and %
    the \MyNoSpellcheck{Tatra Banka} Foundation grant No.~\MyNoSpellcheck{2009sds057}.}%
\and%
  Kate Larson\inst{2}%
\and%
  Ľuboš Steskal\inst{1}%
  \thanks{Supported in part by %
    the National Scholarship Programme of the Slovak Republic, and %
    the \MyNoSpellcheck{Tatra Banka} Foundation grant No.~\MyNoSpellcheck{2009sds056}.}%
}
\institute{
  Dept. of Computer Science, Comenius University, Slovak Republic\\
  \email{\textrm{\{}macko,steskal\textrm{\}}@dcs.fmph.uniba.sk}
\and
  Cheriton School of Computer Science, University of Waterloo, Canada\\
  \email{klarson@cs.uwaterloo.ca}
}
\maketitle

\begin{abstract}
We study the properties of Braess's paradox in the context of the model of
congestion games with flow over time introduced by Koch and Skutella. We
compare them to the well known properties of Braess's paradox for Wardrop's
model of games with static flows. We show that there are networks which do not
admit Braess's paradox in Wardrop's model, but which admit it in the model
with flow over time. Moreover, there is a topology that admits a much
more severe Braess's ratio for this model. Further, despite its symmetry for
games with static flow, we show that Braess's paradox is not symmetric for
flows over time. We illustrate that there are network topologies which exhibit
Braess's paradox, but for which the transpose does not. Finally, we conjecture
a necessary and sufficient condition of existence of Braess's paradox in
a network, and prove the condition of existence of the paradox either in the
network or in its transpose.
\end{abstract}



\section{Introduction} 

Selfish routing and congestion games on networks have been analyzed mainly with
respect to only static flows. The most prevalent model of congestion games with
static flows is Wardrop's model
\cite{Haurie_Marcotte:1985:On_the_relationship_between_Nash-Cournot_and_Wardrop_equilibria,%
Wardrop:1952:Some_theoretical_aspects_of_road_traffic_research} extensively
studied by Roughgarden and Tardos
\cite{Roughgarden:2006:Selfish_routing_and_the_price_of_anarchy,%
Roughgarden_Tardos:2002:How_bad_is_selfish_routing}. A game in Wardrop's
model is played by an infinite set of players each of which is selfishly
routing only a negligible amount of his traffic.

In various applications, e.g.~road traffic control and communication networks,
however, flow variation over time is a crucial feature. Flow congestion on
links and the time to traverse them may change over time in such applications
and the flow does not reach its destination instantaneously, but it travels
through the network at a certain speed determined by link transit times. We
model these phenomena by \emph{flows over time} (also known as \emph{dynamic
flows}) introduced by Ford and Fulkerson
\cite{Ford_Fulkerson:1958:Constructing_maximal_dynamic_flows_from_static_flows,%
Ford_Fulkerson:1962:Flows_in_Networks}.

Nash equilibria for flows over time were introduced by Vickrey
\cite{Vickrey:1969:Congestion_theory_and_transport_investment} and Yagar
\cite{Yagar:1971:Dynamic_traffic_assignment_by_individual_path_minimization_and_queuing}
and mainly studied within the traffic community. For a survey see, e.g.~%
\cite{Peeta_Ziliaskopoulos:2001:Foundations_of_dynamic_traffic_assignment}. In
2009, Koch and Skutella
\cite{Koch_Skutella:2009:Nash_equilibria_and_the_price_of_anarchy_for_flows_over_time},
defined a new variant of flows over time and introduced the notion of the
price of anarchy for them. This model is based on the deterministic queueing
model introduced by Vickrey
\cite{Vickrey:1969:Congestion_theory_and_transport_investment}, in which if at
some point in time, more flow tries to enter a link than its capacity allows,
the flow queues up at the link tail and waits until it may actually enter the
link. The total time spent by a flow particle to traverse a single link is
then the sum of the waiting time in the link queue and the actual time to
traverse the link. In the model with flow over time introduced by Koch and
Skutella, every link of a given network has a fixed capacity and a fixed free flow
transit time. The link capacity bounds the maximal rate at which the flow may
traverse the link and the link free flow transit time expresses the time
a flow particle spends traveling from the link tail to its head.

It is a well known property of selfish routing with static flows that adding
a new link to a network does not necessarily decrease the congestion in Nash
equilibrium, but, paradoxically, it may even increase it, and so increase the
cost of routing through the network. This phenomenon, discovered by Braess
\cite{Braess:1968:Uber_ein_Paradoxon_aus_der_Verkehrsplanung}%
\vphantom{\cite{Braess_Nagurney_Wakolbinger:2005:On_a_paradox_of_traffic_planning}},
is called Braess's paradox. For a survey see,
e.g.~\cite{Roughgarden:2006:Designing_networks_for_selfish_users_is_hard}.

For Wardrop's model of static flows, it is known
\cite{Lin_Roughgarden_Tardos:2004:A_stronger_bound_on_Braess's_Paradox,%
Roughgarden:2006:Designing_networks_for_selfish_users_is_hard} that the ratio
by which the efficiency of a network may improve by removing any number of its
links, i.e.~Braess's ratio, is at most $\lfloor{}n/2\rfloor$, where $n$ is the
number of network nodes. This bound is tight
\cite{Kameda:2002:How_harmful_the_paradox_can_be_in_the_Braess_networks,%
Roughgarden:2006:Designing_networks_for_selfish_users_is_hard} for Wardrop's
model. In this paper, we will prove that this bound does not generalize for the
model of games with flow over time, and that there is a topology which admits a
much more severe Braess's ratio.
For flows over time nothing was known in this respect. Akamatsu and
Heydecker
{\cite{Akamatsu_Heydecker:2003:Detecting_Dynamic_Traffic_Assignment_Capacity_Paradoxes_in_Saturated_Networks}}
considered similar paradoxes for flows over time from a different point of
view.

In principle, the only kind of topology that admits Braess's paradox in static
flows is the Wheatstone network, see
Fig.~\ref{img:FoT-model:paradox_topologies}(a), also known as the $\theta$-network. For Wardrop's model of static
flows, it is known
\cite{Milchtaich:2006:Network_topology_and_the_efficiency_of_equilibrium} that
Braess's paradox may arise on and only on networks which contain the
Wheatstone network as a topological minor.
Recall, that a network $H$ is called a topological minor of a network $G$
if a subdivision of $H$ is isomorphic to a subgraph of $G$.
The networks that do not contain
the Wheatstone network as a topological minor are usually called
series-parallel \cite{Duffin:1965:Topology_of_series-parallel_networks}, as
they can be inductively composed by a number of series and parallel
compositions from an edge. In other words, a network admits Braess's paradox
in Wardrop's model if and only if it is not series-parallel. We will show that
there are networks which do not admit Braess's paradox in Wardrop's model, but
which admit it in the model with flow over time. Thus, we will introduce
a class of series-parallel networks, for which there are instances of the
model with flow over time that admit Braess's paradox. An example of such
a network is shown in Fig.~\ref{img:FoT-model:paradox_topologies}(b).
 Actually, all these networks are even extension-parallel.

\begin{myfigure} \label{img:FoT-model:paradox_topologies} 
  \caption{%
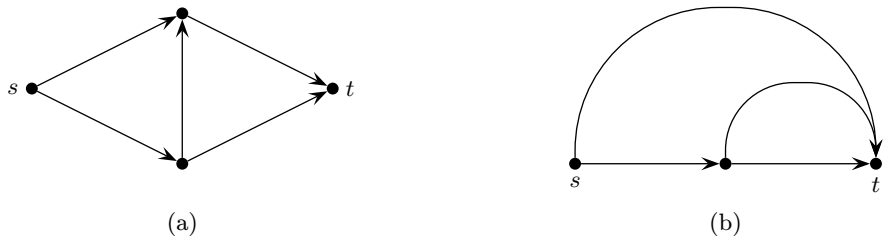
    (a) The Wheatstone network, in principle the only topology that admits Braess's paradox in static flows; %
    (b) The new topology for Braess's paradox in flows over time.%
  }%
  \hbox to \hsize\bgroup%
    \hfill%
    \begin{pspicture}[linewidth=0.5pt,dotscale=1.5,arrowscale=2](-1,-1.1)(5,2.2)%
      \dotnode(0,1){A1}\uput[180](0,1){$\FoTSourceNode$}%
      \dotnode(2,0){A2}%
      \dotnode(2,2){A3}%
      \dotnode(4,1){A4}\uput[0](4,1){$\FoTSinkNode$}%
      \ncline{->}{A1}{A2}%
      \ncline{->}{A1}{A3}%
      \ncline{->}{A2}{A3}%
      \ncline{->}{A2}{A4}%
      \ncline{->}{A3}{A4}%
      \rput(2,-0.8){(a)}%
    \end{pspicture}
    \hfill%
    \begin{pspicture}[linewidth=0.5pt,dotscale=1.5,arrowscale=2](-1,-1.1)(5,2.2)%
      \dotnode(0,0){A1}\uput[270](0,0){$\FoTSourceNode$}%
      \dotnode(2,0){A2}%
      \dotnode(4,0){A3}\uput[270](4,0){$\FoTSinkNode$}%
      \ncline{->}{A1}{A2}%
      \ncline{->}{A2}{A3}%
      \ncdiag[angle=90,arm=2,linearc=1.9]{->}{A1}{A3}%
      \ncdiag[angle=90,arm=1,linearc=0.9]{->}{A2}{A3}%
      \rput(2,-0.8){(b)}%
    \end{pspicture}
    \hfill%
  \egroup%
\end{myfigure} 

As the Wheatstone network is symmetric, or more precisely, as it is isomorphic
to its transpose, a network $\WrdGraph$ admits Braess's paradox in Wardrop's
model if and only if its transpose $\MyGraphTranspose{\WrdGraph}$ admits
Braess's paradox as well. Moreover, we know that an instance of Wardrop's
model on a network $\WrdGraph$ admits Braess's paradox if and only if the
instance on the transpose network $\MyGraphTranspose{\WrdGraph}$ with the same
latency functions and the same flow supply admits it as well.

We will illustrate that in the model with flow over time there exist network
topologies which exhibit Braess's paradox, but for which the transpose does
not. Also, we will show that there is an infinite set of instances of the
model with flow over time, which admit Braess's paradox, but none of the
corresponding instances on their transpose networks with the same traffic
supply and capacities and the same free flow transit times admits it any more.
At the end of this paper, we conjecture a necessary and sufficient condition
of existence of Braess's paradox in a network, and prove the condition of
existence of the paradox either in the network or in its transpose.

Further discussion on the results can also be found in Macko's PhD thesis
{\cite{Macko:2010:The_Price_of_Anarchy_in_Network_Congestion_Games}}.

\bigskip%
The structure of this paper is as follows\colon{} In Section \ref{sec:model},
we provide the formal definition of Koch's and Skutella's model of games with
flow over time and define all notations we use later in this paper. Then, in
Section \ref{sec:lower_bound}, we prove the lower bound on Braess's ratio for
the model of games with flow over time, and we show that there are networks
which admit Braess's paradox in this model, but which do not admit it in
Wardrop's model. In Section \ref{sec:asymmetry}, we show that Braess's paradox
is not symmetric in this model. And finally, in Section
\ref{sec:iff_condition}, we provide the necessary and sufficient conditions of
existence of Braess's paradox in a network.


\section{The Model} 
\label{sec:model}

In this section, following its original definition introduced by Koch and Skutella
\cite{Koch_Skutella:2009:Nash_equilibria_and_the_price_of_anarchy_for_flows_over_time},
we define the model of games with flow over time.

An instance of a game with flow over time is given by a tuple $\FoTInstance$,
where $\FoTGraph$ is a network modeled by a directed graph $\FoTGraph=(V,E)$,
$\FoTSourceNode\in{}V$ and $\FoTSinkNode\in{V}$ are source and sink nodes of
$\FoTGraph$, respectively, $\FoTCapacity=\{\FoTCapacityLink{e}\}_{e\in{}E}$ is
a vector of \emph{link capacities} with all $\FoTCapacityLink{e}>0$,
$\FoTTransitTime=\{\FoTTransitTimeLink{e}\}_{e\in{}E}$ is a vector of
\emph{link free flow transit times} with all $\FoTTransitTimeLink{e}\geq{}0$,
and $\FoTSupply>0$ is an amount of the game input supply. We assume that there
is at least one path from $\FoTSourceNode$ to $\FoTSinkNode$ in $\FoTGraph$.
Let $\FoTPathsST$ denote the set of all $\FoTST$-paths in $\FoTGraph$.
 Note that by the term path we mean a simple path.

The fundamental concept to this model are \emph{waiting queues} that  accumulate
at the tails of network links if more flow wants to traverse a link than
its capacity allows. Therefore, the link capacity bounds its outflow, that is
the rate at which the flow leaves the link. The total \emph{transit time} of
a flow particle through a network link $e$ at time $\theta$ is the sum of
the \emph{waiting time} $\FoTWaitingTimeLink{e}(\theta)$ in the link queue at
time $\theta$ and the link free flow transit time
$\FoTTransitTimeLink{e}$. The link free flow transit time determines the time
the flow particle needs to traverse the link after leaving its waiting queue.
The term \emph{flow particle} represents an infinitesimally small flow unit
that traverses the network along a single path.

For a given link $e$, the \emph{actual link inflow}
$\FoTInflowLink{e}(\theta)$ is a function that determines the flow rate at
which the flow enters the link $e$ at its tail at time $\theta\geq{}0$.
Similarly, the \emph{actual link outflow} $\FoTOutflowLink{e}(\theta)$ is
a function that determines the flow rate at which the flow leaves the link $e$
at its head at time $\theta\geq{}0$. We have
$\FoTInflowLink{e}(\theta)\geq{}0$ and $\FoTOutflowLink{e}(\theta)\geq{}0$ for
all $\theta\geq{}0$. We usually omit the word actual and write just the link
in- and outflow. Further, the \emph{cumulative link inflow}
$\FoTCInflowLink{e}(\theta)$ is a function that determines the total amount of
flow that entered the link $e$ until $\theta\geq{}0$, and the \emph{cumulative
link outflow} $\FoTCOutflowLink{e}(\theta)$ is a function that determines the
total amount of flow that left the link $e$ until $\theta\geq{}0$. Thus,
$\FoTCInflowLink{e}(\theta)=\int_0^{\theta}\FoTInflowLink{e}(\vartheta)\dt[\vartheta]$
and
$\FoTCOutflowLink{e}(\theta)=\int_0^{\theta}\FoTOutflowLink{e}(\vartheta)\dt[\vartheta]$
for all $\theta\geq{}0$. Note that all cumulative in- and outflows are
continuous and nondecreasing. A \emph{flow over time} is a vector
$\FoTFlow=\{(\FoTInflowLink{e},\FoTOutflowLink{e})\}_{e\in{}E}$ of  pairs of the in- and
outflows of all network links.

We say that a flow over time $\FoTFlow$ is feasible if it satisfies the
following conditions. The outflow of every link $e\in{}E$ is upper bounded by
its capacity, therefore
\begin{equation}\label{eqn:model:feasible:capacity}
\FoTOutflowLink{e}(\theta)~\leq~\FoTCapacityLink{e},
\end{equation}%
for all $\theta\geq{}0$. The flow leaves a link $e$ after and only after it
waits in its waiting queue and then it traverses the whole link, so for all
$e\in{}E$ and $\theta\geq{}0$ we have
\begin{equation}\label{eqn:model:feasible:link}
\FoTCInflowLink{e}(\theta)-\FoTCOutflowLink{e}(\theta+\FoTWaitingTimeLink{e}(\theta)+\FoTTransitTimeLink{e})~=~0.
\end{equation}%
All flow that enters a node $v$ continues immediately into node $v$ out-links,
and obviously, only flow that just entered the node $v$ may continue
into its out-links. This condition has two exceptions, namely the source node
and the sink node. The amount of flow that leaves the source node through its
out-links is always larger than the amount of flow that enters it through its
in-links. This difference is exactly the network supply $\FoTSupply$.
Reciprocally, the amount of flow that enters the sink node through its
in-links may be  larger than the amount of flow that leaves it through its
out-links. This difference  at time $\theta\geq{}0$ is called the \emph{actual
sink flow} and expresses the amount of flow that successfully finished its route
from the source to the sink node at time $\theta$. We denote it by
$\FoTSinkFlow(\theta)$, and again, we usually omit the word actual and write
just the sink flow. The \emph{cumulative sink flow} $\FoTCSinkFlow(\theta)$ is
a function that determines the total amount of flow that finished its route
until $\theta\geq{}0$, hence
$\FoTCSinkFlow(\theta)=\int_0^{\theta}\FoTSinkFlow(\vartheta)\dt[\vartheta]$.
Therefore, for all $\theta\geq{}0$ the following condition must hold\colon{}
\begin{equation}\label{eqn:model:feasible:node}
\sum_{e\in\MyGraphInlinksNode{v}}\FoTOutflowLink{e}(\theta)
  ~-~\sum_{e\in\MyGraphOutlinksNode{v}}\FoTInflowLink{e}(\theta)
  ~=~
  \begin{cases}
    ~0                    & \text{ for }v\in{}V\setminus\{\FoTSourceNode,\FoTSinkNode\}, \cr
    ~{-}\FoTSupply        & \text{ if }v=\FoTSourceNode, \cr
    ~\FoTSinkFlow(\theta) & \text{ if }v=\FoTSinkNode,
  \end{cases}
\end{equation}%
where $\MyGraphInlinksNode{v}$ and $\MyGraphOutlinksNode{v}$ are the sets of
all in- and out-links of the node $v$, respectively.

Finally, the waiting time on any link $e$ may not be negative, and if there is
a nonempty waiting queue on the tail of the link $e$, the rate, at which the
flow leaves the queue, must utilise the entire link capacity. If a flow
particle leaves the waiting queue on the link $e$ at time $\theta$, then
by condition (\ref{eqn:model:feasible:link}), it will leave the link $e$
at time $\theta+\FoTTransitTimeLink{e}$. Therefore, if a flow particle
enters the link $e$ at time $\theta$, the cumulative amount of flow that
entered the waiting queue until now is $\FoTCInflowLink{e}(\theta)$ and the
cumulative amount of flow that left the waiting queue until now is
$\FoTCOutflowLink{e}(\theta+\FoTTransitTimeLink{e})$. Hence,
$\FoTCInflowLink{e}(\theta)-\FoTCOutflowLink{e}(\theta+\FoTTransitTimeLink{e})$
is the amount of flow waiting in the queue at time $\theta$. So, the
current waiting time in the queue of the link $e\in{}E$ at time
$\theta\geq{}0$ is
\begin{displaymath}\label{eqn:model:feasible:waiting_time}
\FoTWaitingTimeLink{e}(\theta)
  ~=~\frac{\FoTCInflowLink{e}(\theta)-\FoTCOutflowLink{e}(\theta+\FoTTransitTimeLink{e})}%
           {\FoTCapacityLink{e}},
\end{displaymath}%
and for all $e\in{}E$ and all $\theta\geq{}0$ the following condition must
hold\colon{}
\begin{equation}\label{eqn:model:feasible:queue}
\FoTWaitingTimeLink{e}(\theta)~\geq~0\text{\hskip 1cm and\hskip 1cm}
\FoTWaitingTimeLink{e}(\theta)~>~0
  ~\Rightarrow~\FoTOutflowLink{e}(\theta+\FoTTransitTimeLink{e})~=~\FoTCapacityLink{e}.
\end{equation}%
Note that the function
$\theta\rightarrow\theta+\FoTWaitingTimeLink{e}(\theta)$ is increasing and
continuous. This means that no flow particle may overtake any other flow
particle in the link waiting queue. Similarly, it may not overtake any other
flow particle in the rest of the link as well, as the link free flow transit
time is constant for all flow particles. This means that every link in
a feasible flow is FIFO.

\bigskip%
A game with flow over time is a strategic game, in which every flow particle
is an independent player with an infinitesimally small amount of traffic. The
flows of particular players enter the network at the source node such that the
total amount of flow that entered the network at any point of time
$\theta\geq{}0$ is equal to the game input supply $\FoTSupply$. Every player,
before his piece of flow enters the network, independently chooses his
\emph{strategy}, that is a path from the source to the sink node his flow will
route along. Then his flow enters the network and follows this path as quickly
as possible. For every network \FoTST-path $p$, let
$\FoTPathFlow{p}(\theta)$ denote the amount of flow that entered the network at time
$\theta\geq{}0$ and is going to follow the path $p$. We know that
$\sum_{p\in\FoTPathsST}\FoTPathFlow{p}(\theta)=\FoTSupply$ for all
$\theta\geq{}0$.

Koch and Skutella
\cite{Koch_Skutella:2009:Nash_equilibria_and_the_price_of_anarchy_for_flows_over_time}
defined a feasible flow over time to be a Nash equilibrium, or in other words
to be a Nash flow over time, if and only if the flow is sent only
over currently shortest paths, or equivalently, if and only if no flow
overtakes any other flow.

 For a fixed flow over time, let $\FoTLabelFuncNode{v}(\theta)$
denote the earliest point in time when a flow
particle that entered the network at time $\theta$ may arrive at the node
$v$. Then
\begin{eqnarray*}
\FoTLabelFuncNode{\FoTSourceNode}(\theta)
  ~&=&~\theta\qquad\text{ and }\\
\FoTLabelFuncNode{w}(\theta)
  ~&=&~\min\{\FoTLabelFuncNode{v}(\theta)
     +\FoTWaitingTimeLink{e}(\FoTLabelFuncNode{v}(\theta))
     +\FoTTransitTimeLink{e}~|~e=vw\in\MyGraphInlinksNode{w}\}
\end{eqnarray*}%
for every node $w\in{}V\setminus\{\FoTSourceNode\}$ and all $\theta\geq{}0$.
We call these functions \emph{label functions}. Note that the label functions
are nondecreasing and continuous.

We say that a flow is \emph{sent only over currently shortest paths} if for
every link $e=vw\in{}E$ and all $\theta\geq{}0$ we have
\begin{displaymath}
\FoTLabelFuncNode{w}(\theta)
  ~<~\FoTLabelFuncNode{v}(\theta)
      +\FoTWaitingTimeLink{e}(\FoTLabelFuncNode{v}(\theta))
      +\FoTTransitTimeLink{e}
~\Rightarrow~
  \FoTInflowLink{e}(\FoTLabelFuncNode{v}(\theta))~=~0.
\end{displaymath}%
Similarly, we say that \emph{no flow overtakes any other flow} if, for every
flow particle, the amount of flow that entered the network before this flow
particle equals the amount of flow that left the network before this flow
particle. That is, if
$\FoTSupply\cdot\theta=\FoTCSinkFlow(\FoTLabelFuncNode{t}(\theta))$ for all
$\theta\geq{}0$.

Koch and Skutella showed that flow over time is sent only over currently
shortest paths if and only if no flow overtakes any other flow. This gives us
a pair of handy characterizations of Nash equilibria for flows over time.
Finally, they showed that for every instance of the model of games with flow
over time there exists a flow in a Nash equilibrium.

For convenience in the rest of this paper, let $\FoTPricePath{p}(\theta)$
denote the time spent by a flow particle traveling along a network path $p$ if
the flow particle entered the queue at the first link of $p$ at time
$\theta\geq{}0$. We know that if the path $p$ consists only of one link
$e\in{}E$, then
$\FoTPriceLink{e}(\theta)=\FoTWaitingTimeLink{e}(\theta)+\FoTTransitTimeLink{e}$.
If $p$ contains more links, let $e_1$ denote its first link and $p'$ the rest
of the path $p$, that is $p=e_1p'$. Then
$\FoTPricePath{p}(\theta)=\FoTPricePath{p'}(\theta+\FoTWaitingTimeLink{e_1}(\theta)+\FoTTransitTimeLink{e_1})$.
We call these functions \emph{latency functions} or \emph{latencies}.

Further, let $\FoTPriceNode{v}(\theta)$ denote the shortest time in which a flow
particle may get to a node $v\in{}V$ if the flow particle entered the network
source node at time $\theta\geq{}0$. That is,
$\FoTPriceNode{v}(\theta)=\min_{p\in\MyPathsFromTo{\FoTSourceNode}{v}}\{\FoTPricePath{p}(\theta)\}$. Notice that
$\FoTPriceNode{v}(\theta)=\FoTLabelFuncNode{v}(\theta)-\theta$.

\bigskip%
In this paper we investigate Braess's paradox with respect to the social
cost function $\CngSocialCost$, which expresses the
maximum experienced latency of a flow particle.
For a feasible flow
$\FoTFlow$ of an instance $\FoTInstanceName{A}$ of a game with flow over
time, it is defined as follows\colon{}
\begin{displaymath}
\CngSocialCost(\FoTFlow)
~=~\sup_{\theta\geq0}\max_{p\in\FoTPathsST}
   \Big([\FoTPathFlow{p}(\theta)>0]\cdot\FoTPricePath{p}(\theta)\Big).
\end{displaymath}%
So, we are taking into account the supremum
of the latencies of all paths over the points in time in which
a non-negligible amount of flow used the particular path. By a non-negligible
amount of flow we mean a strictly positive amount of flow.

 For every instance of a game with flow over time, we believe, that all
its Nash flows are in principle equivalent, in the sense that their social costs
are equal. However, this has not yet been proven. So we define the Braess's ratio
with respect to the worst case Nash flow, where by the worst Nash flow we mean the
Nash flow with the highest social cost. Nevertheless, all instances of games
with flow over time we use in our proofs have all their Nash flows
provably equivalent. 

Let $\FoTInstanceName{A}=\FoTInstance$ be an instance of a game with flow over
time on a network $\FoTGraph$ and $\FoTFlow^*$ its  worst Nash flow. We say that
$\CngBraessRatio(\FoTInstanceName{A})$ is Braess's ratio of the instance
$\FoTInstanceName{A}$ with respect to the social cost function
$\CngSocialCost$, and define it as follows\colon{}
\begin{displaymath}
\CngBraessRatio(\FoTInstanceName{A})
~=~\max\left\{
     \frac{\CngSocialCost(\FoTFlow^*)}{\CngSocialCost(\FoTFlow^*_H)}
  ~\middle|~
     H\subseteq\FoTGraph
   \right\},
\end{displaymath}%
where $\FoTFlow^*_H$ is the  worst Nash flow of the instance $\FoTInstance[G=H]$ on
the subgraph $H$. Braess's ratio of  a nonempty class of instances of
games with flow over time is the supremum of Braess's ratios of particular
instances.

We say that an instance admits Braess's paradox if its Braess's ratio is
strictly greater than one. Similarly, we say that a network admits 
Braess's paradox if there is an instance on this network with its Braess's
ratio strictly greater than one.

For simplicity, we will write the value of the social cost
$\CngSocialCost(\FoTFlow^*)$ of  the worst Nash flow $\FoTFlow^*$ of an instance
$\FoTInstanceName{A}$ as $\CngSocialCostNE(\FoTInstanceName{A})$.


\section{Lower Bound on Braess's Ratio} 
\label{sec:lower_bound}

In this section we provide a lower bound on Braess's ratio for the model of games
with flow over time, and we show that there is a topology which admits Braess's
paradox in this model, but which does not  admit it in games with static flows.

Let's consider an instance $\FoTInstanceName{A}_n\assign\FoTInstance[G=M_n]$
of a game with flow over time on the network $M_n$, for $n\geq{}2$, as
shown in Fig.~\ref{fig:new_paradox_topology} with the source node
$\FoTSourceNode=v_1$ and the sink node $\FoTSinkNode=v_n$. The input supply of
the network is $\FoTSupply=\alpha_0$. The free flow transit times and the
capacities of the network links are defined as follows\colon{}
$\FoTTransitTimeLink{e_k}=0$, $\FoTTransitTimeLink{f_k}=T$ and
$\FoTCapacityLink{e_k}=\alpha_k$, for $1\leq{}k\leq{}n-1$,
$\FoTCapacityLink{f_k}=\alpha_{k-1}-\alpha_k$, for $1\leq{}k\leq{}n-2$ and
$\FoTCapacityLink{f_{n-1}}=\alpha_{n-2}$, where $T>0$ and
{$0<\alpha_{n-1}<\dots<\alpha_2<\alpha_1<\alpha_0=\FoTSupply$}.
Let $p_1$ denote the
$\FoTST$-path consisting of the single link $f_1$, $p_k$ the $\FoTST$-path
$e_1e_2\dots{}e_{k-1}f_k$, for $2\leq{}k<n$, consisting of several $e$-links
and the link $f_k$, and finally let $p_0$ denote the path $e_1e_2\dots{}e_{n-1}$
that uses only $e$-links, but no $f$-link. We will show that Braess's
ratio of such an instance may be arbitrarily close to $n-1$, depending only on
$\alpha$'s we choose.

\begin{myfigure} \label{fig:new_paradox_topology} 
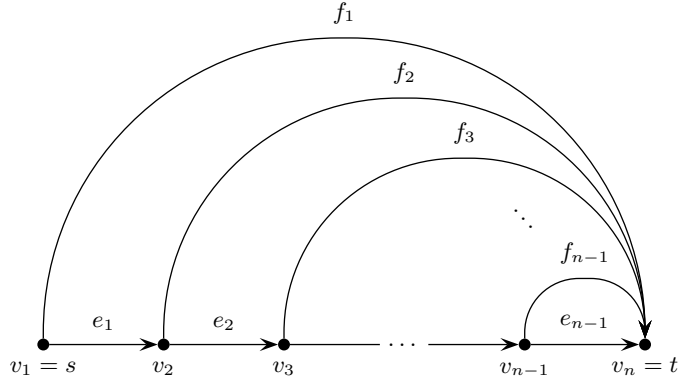
  \caption{The network $M_n$. The network topology which admits a severe Braess's Paradox for %
  congestion games with flows over time, but which does not admit the paradox for Wardrop's model. %
  All these networks are series-parallel, and even extension-parallel.}%
  \begin{pspicture}[linewidth=0.5pt,dotscale=1.5,arrowscale=2,unit=0.8cm](-1,-0.6)(9,4.8)%
    \dotnode(0,0){A1}\uput[270](0,0){$\vphantom{t}v_1=\FoTSourceNode$}%
    \dotnode(2,0){A2}\uput[270](2,0){$\vphantom{t}v_2$}%
    \dotnode(4,0){A3}\uput[270](4,0){$\vphantom{t}v_3$}%
    \dotnode(8,0){Am}\uput[270](8,0){$\vphantom{t}v_{n-1}$}%
    \dotnode(10,0){An}\uput[270](10,0){$\vphantom{t}v_n=\FoTSinkNode$}%
    \ncline{->}{A1}{A2}\naput{$e_1$}%
    \ncline{->}{A2}{A3}\naput{$e_2$}%
    \ncline{->}{A3}{Am}\ncput*{$\dots$}%
    \ncline{->}{Am}{An}\naput{$e_{n-1}$}%
    \ncdiag[angle=90,arm=5,linearc=4.9]{->}{A1}{An}\naput{$f_1$}%
    \ncdiag[angle=90,arm=4,linearc=3.9]{->}{A2}{An}\naput{$f_2$}%
    \ncdiag[angle=90,arm=3,linearc=2.9]{->}{A3}{An}\naput{\rnode[Bl]{Fm}{$f_3$}}%
    \ncdiag[angle=90,arm=1,linearc=0.9]{->}{Am}{An}\naput{\rnode[Bl]{Fn}{$f_{n-1}$}}%
    \ncline[linestyle=none]{Fm}{Fn}\ncput*[nrot=:U,npos=0.66]{$\dots$}%
  \end{pspicture}
\end{myfigure} 

In  every Nash equilibrium at time zero, all first flow particles follow only the
path $p_0$, as it is the only $\FoTST$-path with zero free flow transit time
and so the only $\FoTST$-path with zero  latency. As
$\FoTSupply>\FoTCapacityLink{e_1}>\FoTCapacityLink{e_2}>\dots>\FoTCapacityLink{e_{n-1}}$,
a linearly increasing waiting queue  accumulates on every $e$-link and the
total transit times  on the path $p_0$ and all paths $p_k$ ($k\geq{}2$) linearly
increase with the time when a flow particle entered the network. Since the
queue  accumulates on every $e$-link and the free flow transit times of all
$f$-links are equal, the  latency on every path $p_k$ is strictly greater than
the  latency on the path $p_{k-1}$ at any positive time $\theta$ when a flow
particle entered the network, until  time $\theta_1$ when the  latency on the
path $p_0$ reaches $T$. At this time, the  latency on the path $p_0$ is equal
to the  latency on the path $p_1$. Therefore, any flow particle that enters the
network at  time $\theta_1$ may follow either the path $p_0$ or the path
$p_1$, but not any other path.

After time $\theta_1$, if $n\geq{}3$, the network supply splits between
the paths $p_0$ and $p_1$ in such a way that the  latency on the path $p_1$
begins to increase uniformly with the  latency on the path $p_0$. In particular,
the path $p_0$ will gain the amount of
$\alpha_{n-1}/(\alpha_0-\alpha_1+\alpha_{n-1})$ and the path $p_1$ the amount
of $(\alpha_0-\alpha_1)/(\alpha_0-\alpha_1+\alpha_{n-1})$ portions of the
supply $\FoTSupply$. As
$\FoTSupply\cdot\alpha_{n-1}/(\alpha_0-\alpha_1+\alpha_{n-1})<\alpha_1$, the
 latency on the link $e_1$ decreases and the waiting queue on the link shortens after
 time $\theta_1$. If we are able to choose  $\alpha$'s such that the
link $e_1$ never drains, the link outflow stays constant and equal to
$\alpha_1$ forever, and the flow on the network induced by the nodes $v_2$
to $v_n$ behaves the same way as a flow on the network $M_{n-1}$
of the instance $\FoTInstanceName{A}_{n-1}$ with an input supply equal to
$\alpha_1$.

Indeed, if we choose $\alpha$'s such that none of the $e$-links ever
drains, the  latency on the link $e_{n-1}$ will increase up to $T$, eventually,
with the  latencies on all other $e$-links positive. Therefore, the maximum
experienced transit time  of a flow particle will be strictly greater
than $T$. The next lemma shows that there are $\alpha$'s such that the maximum
experienced transit time of a flow particle in the instance $\FoTInstanceName{A}_n$
is almost $(n-1)\cdot{}T$.

\begin{lemma} \label{lma:main_lema} 
Let $n\geq{}2$, $0<\varepsilon<1/2n$, $j\geq{}1$ and
$\alpha_k=1+\varepsilon^{j+k}$, for $0\leq{}k\leq{}n-1$. In  every Nash
equilibrium, the transit time $\FoTPriceNode{\FoTSinkNode}(\theta)$ of
a flow particle that entered the instance $\FoTInstanceName{A}_n$ at  time
$\theta>T/\varepsilon^{j+n}$ is\colon{}
\begin{displaymath}
\FoTPriceNode{\FoTSinkNode}(\theta)>(1-2n\varepsilon)\cdot(n-1)\cdot{}T.
\end{displaymath}%
\end{lemma} 

\begin{proof} 
We will prove the lemma by induction on $n$. If $n=2$ the network
$M_2$ consists of only two parallel links $e_1$ and $f_1$ with
capacities $\alpha_1$ and $\alpha_0$, respectively.  In every Nash equilibrium, as $\alpha_0>\alpha_1$,
the  time spent traveling through the network $M_2$
 by a flow particle that entered the network at  time
$\theta>T/\varepsilon^{j+2}>T\cdot\frac{1+\varepsilon^{j+1}}{\varepsilon^j-\varepsilon^{j+1}}=T\cdot\frac{\alpha_1}{\alpha_0-\alpha_1}$ is
$\FoTPriceNode{\FoTSinkNode}(\theta)=T>(1-4\varepsilon)\cdot{}T$.

Now, let's assume that $n\geq{}3$, and consider any fixed Nash flow. The cheapest $\FoTST$-path at the beginning
of the game is the path $p_0$, as it is the only $\FoTST$-path with zero free
flow transit time. So, in every Nash flow, all first flow particles follow only this path.
Therefore, the  latency on the path $p_0$ increases linearly with the time when
a flow particle entered the network, until it becomes equal to the  latency on
some other path $p$, when the flow particles
start to follow also the path $p$. Since then, the  latencies on both paths $p_0$
and $p$ grow evenly until they grow up to the  latency on some other path, and so
on. Eventually, the flow particles will start to use all paths $p_1$ to
$p_{n-1}$, since the total capacity of the network after removing any $f$-link
is smaller than the network supply $\FoTSupply$.

For $1\leq{}k\leq{}n-1$, let $\theta_k$ denote the point in time when the first
flow particle which followed the path $p_k$ entered the network. Since the
free flow times of all $f$-links are equal, we know that\colon{}
\begin{displaymath}
\theta_1\leq\theta_2\leq\dots\leq\theta_{n-1}.
\end{displaymath}%
Further, for $1\leq{}k\leq{}n-1$, let $\mu_k$ denote the point in time when the
first flow particle which followed the path $p_k$ entered the link $f_k$.
Clearly $\theta_1=\mu_1$ and $\theta_k\leq{}\mu_k$, for $k\geq{}2$. Then
$\mu_k$ is the point in time when the cumulative amount of flow that entered
the link $e_k$ minus the cumulative amount of flow that left the link
$e_{n-1}$ divided by the link $e_{n-1}$ capacity is exactly the  latency on the
empty link $f_k$. That is the total waiting time on the path
$e_ke_{k+1}\dots{}e_{n-1}$, for a flow particle that enters the link $e_k$ at
time $\mu_k$, is $T$. Therefore
$\mu_k=T\cdot\frac{\alpha_{n-1}}{\alpha_{k-1}-\alpha_{n-1}}$.

If a flow particle enters the network at  time $\theta\in[0,\theta_1)$ it may
follow only the path $p_0$. So, the amount of traffic that enters the link $e_1$
at time $\theta$ is exactly the network supply $\FoTSupply$. Therefore,
the  latency on the link $e_1$ at  time $\theta_1$ is\colon{}
\begin{align*}
\FoTPriceLink{e_1}(\theta_1)
~=~&\theta_1\cdot\frac{\alpha_0-\alpha_1}{\alpha_1}
~=~\mu_1\cdot\frac{\alpha_0-\alpha_1}{\alpha_1}
~=~T\cdot\frac{\alpha_{n-1}}{\alpha_0-\alpha_{n-1}}\cdot\frac{\alpha_0-\alpha_1}{\alpha_1}\displaybreak[1]\\
~=~&T\cdot\left[1-\frac{\alpha_0(\alpha_1-\alpha_{n-1})}{\alpha_1(\alpha_0-\alpha_{n-1})}\right]\displaybreak[1]\\
~=~&T\cdot\left[1-\varepsilon\cdot\frac{1+\varepsilon^j}{1+\varepsilon^{j+1}}\cdot
	\frac{1-\varepsilon^{n-2}}{1-\varepsilon^{n-1}}\right]
 >~T\cdot(1-2\varepsilon).\displaybreak[1]
\end{align*}%
If a flow particle enters the network at  time
$\theta\in[\theta_{k-1},\theta_{k})$, for $2\leq{}k\leq{}n-1$, it may follow
only paths $p_0$ to $p_{k-1}$. Since the capacity of the path $p_1$ is
$\alpha_0-\alpha_1$ and the total capacity of the network induced by paths
$p_0$ and $p_2$ to $p_{k-1}$ is $\alpha_1-\alpha_{k-1}+\alpha_{n-1}$, the
amount of traffic that follow the link $e_1$ at  time $\theta$ is\colon{}
\begin{displaymath}
\gamma_k
~\assign~\alpha_0\cdot\frac{(\alpha_1-\alpha_{k-1}+\alpha_{n-1})}%
    {(\alpha_1-\alpha_{k-1}+\alpha_{n-1})+(\alpha_0-\alpha_1)}
~=~\alpha_0\cdot\frac{\alpha_1-\alpha_{k-1}+\alpha_{n-1}}{\alpha_0-\alpha_{k-1}+\alpha_{n-1}}.
\end{displaymath}%
As $\gamma_k<\alpha_1$ for all $2\leq{}k\leq{}n-1$, the accumulated queue on
the link $e_1$ decreases during the time between $\theta_1$ and
$\theta_{n-1}$.

Finally, if a flow particle enters the network at  time
$\theta\geq\theta_{n-1}$ it may follow all $\FoTST$-paths. Since the total
capacity of the network $M_n$ is equal to the network supply, the
amount of traffic that follow the link $e_1$ at time $\theta$ is
$\gamma_n\assign\alpha_1$, that is exactly the capacity of the link $e_1$.
Therefore the accumulated queue on the link $e_1$ remains constant after
time $\theta_{n-1}$ and
$\FoTPriceLink{e_1}(\theta)=\FoTPriceLink{e_1}(\theta_{n-1})$ for all
$\theta\geq\theta_{n-1}$.

Remark that
$\gamma_{k+1}-\gamma_k<\varepsilon^j\alpha_0(\alpha_{k-1}-\alpha_{n-1})$ for
$2\leq{}k\leq{n-1}$. We show that despite the fact that the  latency on the link
$e_1$ decreases between the time $\theta_1$ and $\theta_{n-1}$, it does not
decrease too much\colon{}
\begin{align*}
\FoTPriceLink{e_1}(\theta_{n-1})
~=~&\FoTPriceLink{e_1}(\theta_1)-
     \sum_{k=2}^{n-1}\frac{\alpha_1-\gamma_k}{\alpha_1}(\theta_k-\theta_{k-1})\displaybreak[1]\\
~>~&\FoTPriceLink{e_1}(\theta_1)-
     \sum_{k=2}^{n-1}\frac{\gamma_{k+1}-\gamma_k}{\alpha_1}\theta_k\displaybreak[1]\\
~\geq~&\FoTPriceLink{e_1}(\theta_1)-\sum_{k=2}^{n-1}\frac{\gamma_{k+1}-\gamma_k}{\alpha_1}\mu_k\displaybreak[1]\\
~>~&\FoTPriceLink{e_1}(\theta_1)-\sum_{k=2}^{n-1}T\cdot
     \frac{\varepsilon^j\alpha_0(\alpha_{k-1}-\alpha_{n-1})}{\alpha_1}\cdot
     \frac{\alpha_{n-1}}{\alpha_{k-1}-\alpha_{n-1}}\displaybreak[1]\\
~>~&\FoTPriceLink{e_1}(\theta_1)-\sum_{k=2}^{n-1}T\cdot{}2\varepsilon^j
~>~T\cdot(1-2n\varepsilon).\displaybreak[1]
\end{align*}%
So, the accumulated queue on the link $e_1$ never drains and its outflow
is always constant and equal to its capacity $\alpha_1$. Therefore,
we can view the flow in the network induced by nodes $v_2$ to $v_n$ as the
instance $\FoTInstanceName{A}_{n-1}$ on the network $M_{n-1}$ with the
input supply $\alpha_1$.

By induction hypothesis we know that  for every time
$\theta>T/\varepsilon^{(j+1)+(n-1)}$ the  time spent traveling along the path
$e_2e_3\dots{}e_{n-1}$  by a flow particle that entered the path at time $\theta$ is\colon{}
\begin{displaymath}
\FoTPricePath{e_2e_3\dots{}e_{n-1}}(\theta)
~>~(1-2(n-1)\varepsilon)\cdot((n-1)-1)\cdot{}T
~>~(1-2n\varepsilon)\cdot(n-2)\cdot{}T.
\end{displaymath}%
As $\theta_{n-1}\leq\mu_{n-1}<T/\varepsilon^{j+n}$, at any time
$\theta>T/\varepsilon^{j+n}$, the  time spent traveling through the network
$M_n$  by a flow particle that entered the network at time $\theta$ is\colon{}
\begin{align*}
\FoTPriceNode{\FoTSinkNode}(\theta)
~=~&\FoTPriceLink{e_1}(\theta)+\FoTPricePath{e_2e_3\dots{}e_{n-1}}(\theta+\FoTPricePath{e_1}(\theta))\\
~=~&\FoTPriceLink{e_1}(\theta_{n-1})+\FoTPricePath{e_2e_3\dots{}e_{n-1}}(\theta+\FoTPricePath{e_1}(\theta))\displaybreak[1]\\
~>~&(1-2n\varepsilon)\cdot{}T+(1-2n\varepsilon)\cdot(n-2)\cdot{}T\\
~=~&(1-2n\varepsilon)\cdot(n-1)\cdot{}T.\tag*{\qed}
\end{align*}%
\end{proof} 

The previous lemma shows that the maximum experienced transit time of a flow
particle, in a Nash equilibrium of $\FoTInstanceName{A}_n$, may be arbitrarily
close to $(n-1)\cdot{}T$, where $n$ is the number of network nodes. The
following theorem shows that there is a subgraph of $M_n$, for which the
maximum experienced transit time of a flow particle in  every Nash equilibrium is
almost $n-1$ times better than in the original graph.

\begin{theorem} \label{thm:lower_bound}
For every $\varepsilon>0$ and $n\geq3$, the network $M_n$ has a subgraph $H$ such that,
for the instance $\FoTInstanceName{A}_n=\FoTInstance[G=M_n]$ we have\colon{}
\begin{displaymath}
\CngSocialCostNE\FoTInstance[G=M_n]
~>~(1-\varepsilon)\cdot(n-1)\cdot \CngSocialCostNE\FoTInstance[G=H].
\end{displaymath}%
\end{theorem} 

\begin{proof} 
Let $H$ be the network $M_n$ with the link $e_{n-1}$ removed. We see
that free flow transit times of all $\FoTST$-paths in $H$ are the same and
equal to $T$. Moreover the total capacity of the network $H$ is equal to the
supply $\FoTSupply=\alpha_0$, therefore  in every Nash flow at any time $\theta\geq{}0$ all flow
supply distributes between the $\FoTST$-paths evenly and no waiting queues
begin to build on any link. So  the network transit time
remains constant and
$\CngSocialCostNE\FoTInstance[G=H]=\FoTPriceNode{\FoTSinkNode}(\theta)=T$, for
all $\theta\geq{}0$.

By Lemma \ref{lma:main_lema} we know that
$\CngSocialCostNE(\FoTInstanceName{A}_n)>(1-2n\varepsilon')\cdot(n-1)\cdot{}T$
for every sufficiently small $\varepsilon'$. So, for
$\varepsilon'\assign\varepsilon/2n$, we have\colon{}
\begin{displaymath}
\CngSocialCostNE(\FoTInstanceName{A}_n)
~>~(1-\varepsilon)\cdot(n-1)\cdot{}T
~=~(1-\varepsilon)\cdot(n-1)\cdot{}\CngSocialCostNE\FoTInstance[G=H].\tag*{\qed}
\end{displaymath}%
\end{proof} 

Therefore, if we choose a sufficiently small $\varepsilon$, Braess's ratio
of the instance $\FoTInstanceName{A}_n$ gets arbitrarily close to $n-1$.

\begin{corollary}
[Lower bound on Braess's ratio]%
\label{cor:lower_bound} %
For every $n\geq{}3$, Braess's ratio of the class $\MyClass{I}_n$
of all instances of the game with flow over time on networks with $n$ nodes is
$\CngBraessRatio(\MyClass{I}_n)\geq{}n-1$.
\end{corollary} 

\begin{corollary}
[A new topology for Braess's paradox]%
\label{cor:compare_to_W-model}%
For every $n\geq{}3$, there is a network with $n$ nodes, which admits 
Braess's paradox in the model of games with flow over time, but which does not
admit it in Wardrop's model. In particular, it is the network $M_n$.
\end{corollary} 

The construction in Lemma \ref{lma:main_lema} also works if we restrict the
model to instances with only integer link capacities. For a given
$\varepsilon$ from the lemma, take the smallest integer $a$ such that
$1/2^a\leq\varepsilon$, and let $\alpha_k=2^{a(n+j)}+2^{a(n-k)}$. By a proof
similar to the proof of Lemma \ref{lma:main_lema}, we can show that in
a Nash equilibrium the transit time of a flow particle that entered the
network at  time $\theta>T\cdot{}2^{a(j+n)}$ is more than
$(1-n/2^{a-1})\cdot(n-1)\cdot{}T$.

If we restrict the model only to instances with unit link capacities, the
lower bound on Braess's ratio as a function of the number of network nodes
still holds. We only need to replace every network link with integer capacity
$c$ by a set of $c$ parallel links with unit capacities. However, in this case,
the number of network links grows exponentially with $n$ and polynomially with
$1/\varepsilon$.


\section{Asymmetry of Braess's Paradox} 
\label{sec:asymmetry}

For every $n\geq{}3$, we have shown that the instance $\FoTInstanceName{A}_n$
 as defined in the previous section has Braess's ratio arbitrarily close to
$n-1$ for sufficiently small $\varepsilon$, and so it admits Braess's paradox.
Now, we will show that the instance on the transpose network with the same
traffic supply and the same link capacities and free flow times has Braess's
ratio equal to $1$.

\begin{myfigure} \label{fig:Gn_transpose} 
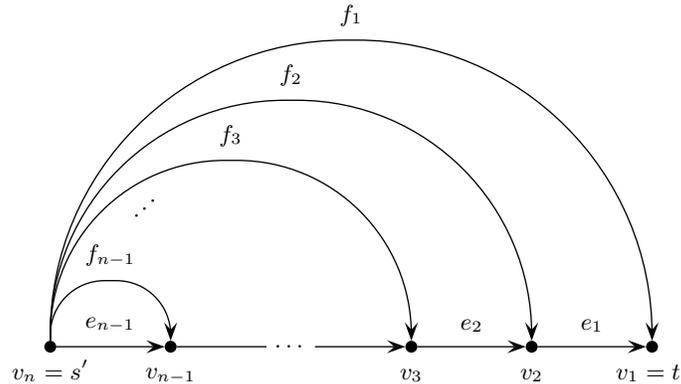
  \caption{The transpose $\MyGraphTranspose{M_n}$ of the network $M_n$, which %
    admits no Braess's paradox in the model of games with flow over time.}%
  \begin{pspicture}[linewidth=0.5pt,dotscale=1.5,arrowscale=2,unit=0.8cm](-1,-0.6)(9,4.8)%
    \dotnode(10,0){A1}\uput[270](10,0){$\vphantom{t'}v_1=\FoTSinkNode'$}%
    \dotnode(8,0){A2}\uput[270](8,0){$\vphantom{t'}v_2$}%
    \dotnode(6,0){A3}\uput[270](6,0){$\vphantom{t'}v_3$}%
    \dotnode(2,0){Am}\uput[270](2,0){$\vphantom{t'}v_{n-1}$}%
    \dotnode(0,0){An}\uput[270](0,0){$\vphantom{t'}v_n=\FoTSourceNode'$}%
    \ncline{<-}{A1}{A2}\nbput{$e_1$}%
    \ncline{<-}{A2}{A3}\nbput{$e_2$}%
    \ncline{<-}{A3}{Am}\ncput*{$\dots$}%
    \ncline{<-}{Am}{An}\nbput{$e_{n-1}$}%
    \ncdiag[angle=90,arm=5,linearc=4.9]{<-}{A1}{An}\nbput{$f_1$}%
    \ncdiag[angle=90,arm=4,linearc=3.9]{<-}{A2}{An}\nbput{$f_2$}%
    \ncdiag[angle=90,arm=3,linearc=2.9]{<-}{A3}{An}\nbput{\rnode[Bl]{Fm}{$f_3$}}%
    \ncdiag[angle=90,arm=1,linearc=0.9]{<-}{Am}{An}\nbput{\rnode[Bl]{Fn}{$f_{n-1}$}}%
    \ncline[linestyle=none]{Fm}{Fn}\ncput*[nrot=:U,npos=0.66]{$\dots$}%
  \end{pspicture}
\end{myfigure} 

Consider an instance
$\MyGraphTranspose{\FoTInstanceName{A}_n}\assign\FoTInstance[G=\MyGraphTranspose{M_n},s=\FoTSourceNode',t=\FoTSinkNode']$
on the network $\MyGraphTranspose{M_n}$, for $n\geq{}2$, as shown in
Fig.~\ref{fig:Gn_transpose}, with an input supply
$d=\alpha_0$. The network $\MyGraphTranspose{M_n}$ is a transpose of the
network $M_n$, that is the network $M_n$ with all its links reversed and its
source and sink nodes swapped. Therefore, the $\MyGraphTranspose{M_n}$ source
and sink nodes are $\FoTSourceNode'=v_n$ and $\FoTSinkNode'=v_1$,
respectively, and the free flow transit times and the capacities of its links
are defined as follows\colon{} $\FoTTransitTimeLink{e_k}=0$,
$\FoTTransitTimeLink{f_k}=T$ and $\FoTCapacityLink{e_k}=\alpha_k$, for
$1\leq{}k\leq{}n-1$, $\FoTCapacityLink{f_k}=\alpha_{k-1}-\alpha_k$, for
$1\leq{}k\leq{}n-2$ and $\FoTCapacityLink{f_{n-1}}=\alpha_{n-2}$, where $T>0$
and {$0<\alpha_{n-1}<\dots<\alpha_2<\alpha_1<\alpha_0=d$}.
Similarly, denote
$\MyGraphTranspose{p_k}$ as the reverse of the path $p_k$. Therefore,
$\MyGraphTranspose{p_0}$ is the path $e_{n-1}e_{n-2}\dots{}e_1$ and
$\MyGraphTranspose{p_k}$ is the path $f_ke_{k-1}e_{k-2}\dots{}e_1$, for
$k\geq{}1$.

We will show, that for all $n\geq{}2$ the instance
$\MyGraphTranspose{\FoTInstanceName{A}_n}$ does not admit Braess's paradox.
That is, there is no subgraph $H$ of the network $\MyGraphTranspose{M_n}$, for
which the maximum experienced transit time of a flow particle in  any Nash
equilibrium of the instance
$\FoTInstance[G=H,s=\FoTSourceNode',t=\FoTSinkNode']$ would be smaller than
the maximum experienced transit time of a flow particle in  any Nash equilibrium
of the instance $\MyGraphTranspose{\FoTInstanceName{A}_n}$.

\begin{lemma} \label{lma:transposed_lema} 
For every $n\geq{}2$, Braess's ratio of the instance
$\MyGraphTranspose{\FoTInstanceName{A}_n}$ is
$\CngBraessRatio(\MyGraphTranspose{\FoTInstanceName{A}_n})=1$.
\end{lemma} 

\begin{proof} 
At the beginning of the game $\MyGraphTranspose{\FoTInstanceName{A}_n}$, in
 every Nash equilibrium, all first flow particles follow only the path
$\MyGraphTranspose{p_0}$, as it is the only cheapest $\FoTST$-path in the
network $\MyGraphTranspose{M_n}$. As the capacity $\alpha_{n-1}$ of
the first link on the path $\MyGraphTranspose{p_0}$, the link $e_{n-1}$, is
strictly smaller than the network supply $\alpha_0$ and smaller than the
capacities of all other $e$-links on the path, it is the only link on which
a queue begins to  accumulate. Therefore, the  latencies on all other links remain
constant. The queue on the link $e_{n-1}$ will grow until the  latency on this
link become equal to $T$. At this point of time, the  latencies on all $\FoTST$-paths will
become the same and equal to $T$. So, the traffic will start to split evenly
among all $\FoTST$-paths and no queues will grow any more, as the network
supply is not larger than the total capacity of the network
$\MyGraphTranspose{M_n}$. Therefore, the maximum experienced transit
time of a flow particle in  every Nash equilibrium of the instance
$\MyGraphTranspose{\FoTInstanceName{A}_n}$ is $T$. So,
$\CngSocialCostNE(\MyGraphTranspose{\FoTInstanceName{A}_n})=T$.

Now, take a subgraph $H$ of the network $\MyGraphTranspose{M_n}$ and
let
$\FoTInstanceName{H}\assign\FoTInstance[G=H,s=\FoTSourceNode',t=\FoTSinkNode']$ denote
the instance of the game on the network $H$. If
$H=\MyGraphTranspose{M_n}$, then  Nash equilibria of both instances
are the same and
$\CngSocialCostNE(\FoTInstanceName{H})=\CngSocialCostNE(\MyGraphTranspose{\FoTInstanceName{A}_n})$.
So, assume $H$ is a proper subset of $\MyGraphTranspose{M_n}$. If $H$
misses any other link then $e_{n-1}$, it is easy to check, that its total
capacity is strictly less than the network supply $\alpha_0$ and so
$\CngSocialCostNE(\FoTInstanceName{H})$ is unbounded. If $H$ contains all links
except $e_{n-1}$ the total capacity of the network is equal to the supply
$\FoTSupply=\alpha_0$ and the free flow transit times of all $\FoTST$-paths in
$H$ are equal to $T$. Therefore, in  every Nash equilibrium, the traffic splits
evenly among all $\FoTST$-paths since the beginning of the game and no waiting
queues build on any link. Thus,  the network transit time is
constant and equal to $T$ forever. So, $\CngSocialCostNE(\FoTInstanceName{H})=T$
and
$\CngBraessRatio(\MyGraphTranspose{\FoTInstanceName{A}_n})=1$.\qed
\end{proof} 

We have proved, that none of the instances
$\MyGraphTranspose{\FoTInstanceName{A}_n}$ admits Braess's paradox. In
fact, it is possible to show that there is no Braess's paradox even for the
instance $\FoTInstance[G=\MyGraphTranspose{M_n},s=\FoTSourceNode',t=\FoTSinkNode']$ with
any $\FoTSupply\leq\alpha_0$, not just $\FoTSupply=\alpha_0$.

By Theorem \ref{thm:lower_bound}, for every $n\geq{}3$, we know that 
Braess's ratio of the instance $\FoTInstanceName{A}_n$ is arbitrarily close to
$n-1$, and so the instance $\FoTInstanceName{A}_n$ admits a rather severe
Braess's paradox. Thus, together with the previous lemma, we get the following
theorem\colon{}

\begin{theorem}
[Braess's paradox asymmetry]%
\label{thm:asymmetry}%
For every $n\geq{}3$, there is an instance of
a game with flow over time on a network $G$ with $n$ nodes, which admits 
Braess's paradox, but for which the corresponding instance
on the transpose network $\MyGraphTranspose{G}$ does not admit it.
In particular, it is the instance $\FoTInstanceName{A}_n$.
\end{theorem} 

So, there is an instance on the network $M_3$, see
Fig.~\ref{fig:G_3_and_company}(a), namely $\FoTInstanceName{A}_3$, that admits
Braess's paradox, and we have shown that the instance
$\MyGraphTranspose{\FoTInstanceName{A}_3}$ on the transpose network
$\MyGraphTranspose{M_3}$, see Fig.~\ref{fig:G_3_and_company}(b), with the same
flow supply and the same link capacities and free flow times does not admit
it. In fact, for this particular transpose network, there is no instance which
would have Braess's ratio strictly larger than $1$, and so, which would admit
Braess's paradox.

\begin{myfigure} \label{fig:G_3_and_company} 
  \caption{%
    (a) $M_3$; %
    (b) $\MyGraphTranspose{M_3}$; %
    (c) $M_3'$; %
    (d) $M_3''$.%
  }%
  \hbox to \hsize\bgroup%
    \hfill%
    \begin{pspicture}[linewidth=0.5pt,dotscale=1.5,arrowscale=2,unit=0.75cm](0,-1)(3,2.3)%
      \dotnode(0,0){A1}\uput[270](0,0){\vphantom{\FoTSinkNode}\FoTSourceNode}%
      \dotnode(2,0){A2}%
      \dotnode(4,0){A3}\uput[270](4,0){\vphantom{\FoTSinkNode}\FoTSinkNode}%
      \ncline{->}{A1}{A2}\naput{$e_1$}%
      \ncline{->}{A2}{A3}\naput{$e_2$}%
      \ncdiag[angle=90,arm=2,linearc=1.9]{->}{A1}{A3}\naput{$f_1$}%
      \ncdiag[angle=90,arm=1,linearc=0.9]{->}{A2}{A3}\naput[npos=1.3]{$f_2$}%
      \rput(2,-1){(a)}%
    \end{pspicture}%
    \hfill%
    \begin{pspicture}[linewidth=0.5pt,dotscale=1.5,arrowscale=2,unit=0.75cm](0,-1)(3,2.3)%
      \dotnode(0,0){A3}\uput[270](0,0){\vphantom{\FoTSinkNode}\FoTSourceNode}%
      \dotnode(2,0){A2}%
      \dotnode(4,0){A1}\uput[270](4,0){\vphantom{\FoTSinkNode}\FoTSinkNode}%
      \ncline{<-}{A1}{A2}\nbput{$e_1$}%
      \ncline{<-}{A2}{A3}\nbput{$e_2$}%
      \ncdiag[angle=90,arm=2,linearc=1.9]{<-}{A1}{A3}\nbput{$f_1$}%
      \ncdiag[angle=90,arm=1,linearc=0.9]{<-}{A2}{A3}\nbput[npos=1.3]{$f_2$}%
      \rput(2,-1){(b)}%
    \end{pspicture}%
    \hfill%
    \begin{pspicture}[linewidth=0.5pt,dotscale=1.5,arrowscale=2,unit=0.75cm](0,-1)(3,2.3)%
      \SpecialCoor%
      \dotnode(0.00,0){A1}\uput[270](A1){\vphantom{\FoTSinkNode}\FoTSourceNode}%
      \dotnode(1.33,0){A2}%
      \dotnode(2.66,0){B1}%
      \dotnode(4.00,0){A3}\uput[270](A3){\vphantom{\FoTSinkNode}\FoTSinkNode}%
      \ncline{->}{A1}{A2}\naput{$e_1$}%
      \ncline{->}{A2}{B1}\naput{$e_2$}%
      \ncline{->}{B1}{A3}\naput{$g$}%
      \myncarc[arcangle=90]{->}{A1}{A3}\naput{$f_1$}%
      \myncarc[arcangle=90]{->}{A2}{B1}\naput{$f_2$}%
      \rput(2,-1){(c)}%
    \end{pspicture}%
    \hfill%
    \begin{pspicture}[linewidth=0.5pt,dotscale=1.5,arrowscale=2,unit=0.75cm](0,-1)(3,2.3)%
      \SpecialCoor%
      \dotnode(0,0){A1}\uput[270](A1){\vphantom{\FoTSinkNode}\FoTSourceNode}%
      \dotnode(2,0){A2}%
      \dotnode(4,0){A3}\uput[270](A3){\vphantom{\FoTSinkNode}\FoTSinkNode}%
      \ncline{->}{A1}{A2}\naput{$e_1$}%
      \ncline{->}{A2}{A3}\naput{$e_2$}%
      \myncarc[arcangle=90,linestyle=none]{A1}{A3}\lput(0.67){\dotnode{B1}}%
      \myncarc[arcangle=60]{->}{A1}{B1}\naput{$f_1$}%
      \myncarc[arcangle=30]{->}{B1}{A3}\naput{$g$}%
      \ncline{->}{A2}{B1}\naput{$f_2$}%
      \rput(2,-1){(d)}%
    \end{pspicture}%
    \hfill%
  \egroup%
\end{myfigure} 

\begin{theorem}[Braess's paradox asymmetry for networks] \label{thm:asymmetry_G3}
There is a network $\FoTGraph$ with an instance of the game with flow over
time that admits Braess's paradox, for which there is no instance of the
game with flow over time on its transpose {$\MyGraphTranspose{\FoTGraph}$}
that would admit the paradox.
 In particular, it is the network {$M_3$}.
\end{theorem} 

%
%
%
%

\begin{proof} 
Let $\FoTGraph=M_3$, by theorem \ref{thm:lower_bound}, we know that
the instance $\FoTInstanceName{A}_3$ on the network $M_3$ has the
Braess's ratio arbitrarily close to $2$ for sufficiently small $\varepsilon$,
hence, it admits the Braess's paradox.

Now, take some instance
$\FoTInstanceName{B}=\FoTInstance[G=\MyGraphTranspose{M_3}]$ on the transpose
$\MyGraphTranspose{M_3}$ of the network $M_3$. If the network supply
$\FoTSupply$ is larger than the total capacity of the union of all paths from
the network source node $\FoTSourceNode$ to the network sink node
$\FoTSinkNode$, then the maximum experienced transit time of a flow particle
in every Nash equilibrium is unbounded. As no subgraph of the network
$\MyGraphTranspose{M_3}$ has its total capacity larger than the total capacity
of the entire network $\MyGraphTranspose{M_3}$, the maximum experienced
transit time of a flow particle in every Nash equilibrium of every
$\MyGraphTranspose{M_3}$ subgraph is unbounded as well. Therefore the instance
$\FoTInstanceName{B}$ does not admit the Braess's paradox.

Thus, let's assume that the network supply $\FoTSupply$ is at most the total capacity of
the union of all paths from the network source node $\FoTSourceNode$ to the
network sink node $\FoTSinkNode$.
If the instance has the network source and
sink nodes chosen differently than in Figure
\ref{fig:G_3_and_company}(b), then there is no path from
$\FoTSourceNode$ to $\FoTSinkNode$ in $\MyGraphTranspose{M_3}$, or the
set of all $\FoTST$-paths forms a subgraph with only one or two parallel links
from $\FoTSourceNode$ to $\FoTSinkNode$. In either case, the instance does not
admit the Braess's paradox, trivially.

So, let's assume that the source and sink nodes of the instance
$\FoTInstanceName{B}$ are chosen as shown in Figure
\ref{fig:G_3_and_company}(b), and consider any fixed Nash equilibrium of the instance $\FoTInstanceName{B}$.
Denote $p_1$ the path $e_2e_1$, $p_2$ the
path $f_2e_1$, and $p_3$ the path $f_1$, where $e_1$, $e_2$, $f_1$ and $f_2$
are the network links as shown in the Figure. For $0\leq{}k\leq{}2$, denote
$t_k$ the time when the first flow particle may enter the path $p_k$ in the
Nash equilibrium, that is the time when the path
$p_k$ becomes one of the shortest $\FoTST$-paths. Note that, in general, even
if a path becomes one of the shortest $\FoTST$-paths, the flow particles may
but do not have to use it at all if there is a sufficient capacity on the
remaining shortest $\FoTST$-paths. If the path $p_k$ never becomes one of the
shortest $\FoTST$-paths in the Nash equilibrium, we say that $t_k=\infty$. Also
note that after a path becomes one of the shortest paths, it remains among the
shortest paths forever.

By checking out all possibilities on relative order of the times $t_k$, we will
show that the instance $\FoTInstanceName{B}$ may not admit the Braess's
paradox. So, take a subgraph $H$ of the network
$\MyGraphTranspose{M_3}$. We will show that the maximum experienced
transit time $\CngSocialCost(\FoTInstanceName{B'})$ of a traffic particle in
every Nash equilibrium of the instance $\FoTInstanceName{B'}=\FoTInstance[G=H]$
restricted to $H$ is at least the maximum experienced transit time
$\CngSocialCost(\FoTInstanceName{B})$ of a traffic particle in the chosen Nash
equilibrium of the instance $\FoTInstanceName{B}$ on the original network
$\MyGraphTranspose{M_3}$\colon{}
\begin{itemize}

\item%
If $t_1\leq{}t_2\leq{}t_3<\infty$ then the free flow transit time of the path
$p_1$ is not bigger than the free flow transit time of the path $p_2$, which
is not bigger than the free flow transit time of the path $p_3$, that is
$\FoTTransitTimeLink{e_2}\leq\FoTTransitTimeLink{f_2}$ and
$\FoTTransitTimeLink{f_2}+\FoTTransitTimeLink{e_1}\leq\FoTTransitTimeLink{f_1}$.
As $t_3<\infty$, the path $p_3$ becomes one of the shortest $\FoTST$-paths at
the time $t_3$, therefore the transit time of every flow particle that enters
the network at this time is exactly $\FoTTransitTimeLink{f_1}$. Since both
$t_1$ and $t_2$ are at most $t_3$, the paths $p_1$ and $p_2$ are the shortest
$\FoTST$-paths at the time $t_3$ as well. Therefore, all network paths are used
since $t_3$, and their total capacity is sufficient for the network supply
$\FoTSupply$. So, the maximum experienced transit time of a flow particle in
the Nash equilibrium is
$\CngSocialCost(\FoTInstanceName{B})=\FoTTransitTimeLink{f_1}$. Now, we have
three cases\colon{}
\begin{itemize}
\item%
If $t_2<t_3$ then the total capacity of the paths $p_1$ and $p_2$ is strictly
smaller than the network supply $\FoTSupply$, as their transit times grow
until $t_3$, when the transit times of all three paths become equal.
Therefore, if $p_3\not\subseteq{}H$ the maximum experienced transit time
$\CngSocialCost(\FoTInstanceName{B'})$ of a flow particle in the Nash
equilibrium of $\FoTInstanceName{B'}$ is unbounded. However, if
$p_3\subseteq{}H$ the path $p_3$ is used in the equilibrium, and the maximum
experienced transit time $\CngSocialCost(\FoTInstanceName{B'})$ of a flow
particle is at least $\FoTTransitTimeLink{f_1}$.
\item%
If $t_1<t_2$ and $t_2=t_3$ then the capacity of the path $p_1$ is strictly
smaller than $\FoTSupply$. Therefore, if neither $p_2\subseteq{}H$ nor
$p_3\subseteq{}H$, then $\CngSocialCost(\FoTInstanceName{B'})$ is unbounded.
Otherwise, if $p_2\subseteq{}H$ or $p_3\subseteq{}H$, then at least one of
these paths is used by the flow, and so
\begin{math}
\CngSocialCost(\FoTInstanceName{B'})\geq\FoTTransitTimeLink{f_2}
  +\FoTTransitTimeLink{e_1}=\FoTTransitTimeLink{f_1}.
\end{math}%
\item%
If $t_1=t_2$ and $t_2=t_3$ then the free flow transit times of all three paths
are equal and
\begin{math}
\CngSocialCost(\FoTInstanceName{B'})
  =\FoTTransitTimeLink{e_2}+\FoTTransitTimeLink{e_1}
  =\FoTTransitTimeLink{f_2}+\FoTTransitTimeLink{e_1}
  =\FoTTransitTimeLink{f_1}
\end{math}%
or is unbounded, depending on the total capacity of $H$.
\end{itemize}

\item%
If $\max\{t_1,t_3\}<t_2<\infty$ then the total capacity of the paths $p_1$ and
$p_3$ is strictly smaller than $\FoTSupply$. If $\FoTCapacityLink{e_2}$ would
not be smaller then $\FoTCapacityLink{e_1}$, the total capacity of the entire
network $\MyGraphTranspose{M_3}$ would be smaller than $\FoTSupply$,
therefore $\FoTCapacityLink{e_2}<\FoTCapacityLink{e_1}$. As
$\FoTCapacityLink{e_2}<\FoTCapacityLink{e_1}$ and the link $f_2$ belongs only
to the path $p_2$, the link $e_1$ does not accumulate any waiting queue and
its transit time remains $\FoTTransitTimeLink{e_1}$ forever. Therefore, the
transit time of the path $p_2$ at the time $t_2$, when it becomes one of the
shortest paths, is exactly
$\FoTTransitTimeLink{f_2}+\FoTTransitTimeLink{e_1}$. Hence,
$\CngSocialCost(\FoTInstanceName{B})=\FoTTransitTimeLink{f_2}+\FoTTransitTimeLink{e_1}$.

Now, if $p_2\not\subseteq{}H$, then $\CngSocialCost(\FoTInstanceName{B'})$ is
unbounded, since the total capacity of the paths $p_1$ and $p_3$ is smaller
than $\FoTSupply$. If $p_2\subseteq{}H$, then
$\CngSocialCost(\FoTInstanceName{B'})$ is at least the free flow transit time
of the path $p_2$, that is
$\FoTTransitTimeLink{f_2}+\FoTTransitTimeLink{e_1}$.

\item%
If $t_3<t_1=t_2<\infty$ then
$\FoTTransitTimeLink{f_1}<\FoTTransitTimeLink{f_2}+\FoTTransitTimeLink{e_1}=\FoTTransitTimeLink{e_2}+\FoTTransitTimeLink{e_1}$
and $\FoTCapacityLink{f_1}<\FoTSupply$. Therefore,
$\CngSocialCost(\FoTInstanceName{B})=\FoTTransitTimeLink{e_2}+\FoTTransitTimeLink{e_1}$.
Now, if $p_1\subseteq{}H$ or $p_2\subseteq{}H$, then
$\CngSocialCost(\FoTInstanceName{B'})\geq\FoTTransitTimeLink{e_2}+\FoTTransitTimeLink{e_1}$.
Otherwise, $\CngSocialCost(\FoTInstanceName{B'})$ is unbounded.

\item%
If $t_1\leq{}t_2<t_3=\infty$ then
$\FoTTransitTimeLink{e_2}\leq\FoTTransitTimeLink{f_2}$ and
$\FoTTransitTimeLink{f_2}+\FoTTransitTimeLink{e_1}<\FoTTransitTimeLink{f_1}$.
As $t_3=\infty$, the path $p_3$ never becomes the shortest path, and so is
never used. Hence, $\FoTSupply\leq\FoTCapacityLink{e_1}$ and no queue
accumulates on the link $e_1$. Therefore,
$\CngSocialCost(\FoTInstanceName{B})=\FoTTransitTimeLink{f_2}+\FoTTransitTimeLink{e_1}$.

Now, if the remainder of the paths $p_1$ and $p_2$ in $H$ has its total
capacity at least $\FoTSupply$, then
$\CngSocialCost(\FoTInstanceName{B'})=\FoTTransitTimeLink{f_2}+\FoTTransitTimeLink{e_1}$,
Otherwise, the path $p_3$ is used, provided that it belongs to $H$. In this
case
$\CngSocialCost(\FoTInstanceName{B'})\geq\FoTTransitTimeLink{f_1}>\FoTTransitTimeLink{f_2}+\FoTTransitTimeLink{e_1}$.
If neither the remainder of the paths $p_1$ and $p_2$ in $H$ has sufficient
capacity nor $p_3\subseteq{}H$, then $\CngSocialCost(\FoTInstanceName{B'})$ is
unbounded.

\item%
The cases if $t_1\leq{}t_3<t_2=\infty$ and if $t_3\leq{}t_1<t_2=\infty$ are
analogous to the previous case.

\item%
If $t_1<t_2=t_3=\infty$ then
$\FoTTransitTimeLink{e_2}<\FoTTransitTimeLink{f_2}$ and
$\FoTTransitTimeLink{e_2}+\FoTTransitTimeLink{e_1}<\FoTTransitTimeLink{f_1}$.
Moreover, the paths $p_2$ and $p_3$ are never used and the path $p_1$ has
enough capacity to support the entire input supply $\FoTSupply$. Hence,
$\CngSocialCost(\FoTInstanceName{B})=\FoTTransitTimeLink{e_2}+\FoTTransitTimeLink{e_1}$.
Now, if $p_1\subseteq{}H$, then
$\CngSocialCost(\FoTInstanceName{B'})=\FoTTransitTimeLink{e_2}+\FoTTransitTimeLink{e_1}$.
Otherwise, the flow particles in $H$ must use some of the remaining paths. In
this case,
$\CngSocialCost(\FoTInstanceName{B'})\geq\min\{\FoTTransitTimeLink{f_1},\FoTTransitTimeLink{f_2}+\FoTTransitTimeLink{e_1}\}>\FoTTransitTimeLink{e_2}+\FoTTransitTimeLink{e_1}$.

\item%
The case if $t_3<t_1=t_2=\infty$ is analogous to the previous case.

\item%
Finally, the cases if $t_2<t_1$ are symmetric to the cases if $t_1<t_2$.
\qed
\end{itemize}
\end{proof} 


\section{Necessary and Sufficient Condition for Braess's Paradox} 
\label{sec:iff_condition}

In this section, we would like to answer the question, which topologies in
general admit Braess's paradox in the model of games with flow over time. So,
we would like to characterize the class of all such networks in this model.
Foremost, we show that every network which contains either the network $M_3$,
or its variations, the network $M_3'$ (see Fig.~\ref{fig:G_3_and_company}(c))
 or the network $M_3''$ (see Fig.~\ref{fig:G_3_and_company}(d)) as
a topological minor admits Braess's paradox in this model, and then we
conjecture that these three networks are essentially the only topologies that
admit Braess's paradox in this model in general.

The networks $M_3'$ and $M_3''$ are very similar to
$M_3$. If we set the free flow transit times and the link capacities
in these two networks the same way as in the network $M_3$ with the
only difference that $\FoTTransitTimeLink{g}=0$ and
$\FoTCapacityLink{g}=\FoTSupply$, where $\FoTSupply$ is the network supply,
the instances on the networks $M_3'$, $M_3''$ and
$M_3$ act the same way, and their social costs and Braess's ratios are
equal. Therefore, both $M_3'$ and $M_3''$ admit Braess's
paradox, since $M_3$ admits it. Notice that despite their similarity,
the networks $M_3$, $M_3'$ and $M_3''$ are not
topological minors of each other. The conjecture that these three networks are
essentially the only topologies that admit Braess's paradox is motivated
by the result, we investigate at the end of this section, that these three
networks together with the network $\MyGraphTranspose{M_3}$ are the
only topologies that admit Braess's paradox if we use the network both
ways, the forward and also the reverse.

\begin{theorem}
[Sufficient condition for Braess's paradox]%
\label{thm:sufficient_cond} %
If a network $\FoTGraph$ contains either the network $M_3$,
$M_3'$  or $M_3''$ as a topological minor, then it admits 
Braess's paradox in the model of games with flow over time.
\end{theorem} 

\begin{proof} 
If the network $\FoTGraph$ contains $M_3$ as a topological minor then
the network $\FoTGraph$ contains a subgraph that can be constructed from
$M_3$ by a number of link subdivisions. Therefore, there are nodes
$v_1$, $v_2$ and $v_3$ in $\FoTGraph$ connected to each other by four
independent paths, as shown in Fig.~\ref{fig:G_3_subdivision}. We will call
them \emph{subdivided paths} and their links \emph{subdivided links}. Take one
link of each subdivided path and denote it by $e_1$, $e_2$, $f_1$ or $f_2$,
respectively, as shown in the figure.

\begin{myfigure} \label{fig:G_3_subdivision} 
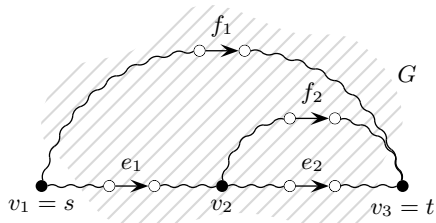
  \caption{$M_3$ as a topological minor of a network $\FoTGraph$.}%
  \begin{pspicture}[linewidth=0.5pt,dotscale=1.5,arrowscale=2,unit=0.6cm](0,-0.4)(6,2.6)%
    \SpecialCoor%
    \definecolor{fill}{rgb}{0.85,0.85,0.85}%
    \psccurve[linestyle=none,fillstyle=hlines,hatchangle=45,hatchcolor=fill]%
      (1.5,0.5)(2.6,-0.8)(5.4,-0.6)(8.22,-0.7)(8.5,0.4)(9.0,1.5)%
      (8.0,3.2)(6.6,3.4)(4.5,4.0)(3.0,3.0)(1.5,2.5)(1.0,1.2)%
    \rput(9.1,2.5){$\FoTGraph$}%
    \dotnode(1,0){A1}\uput[270](A1){$v_1=\FoTSourceNode$}%
    \dotnode(5,0){A2}\uput[270](A2){$v_2$}%
    \dotnode(9,0){A3}\uput[270](A3){$v_3=\FoTSinkNode$}%
    \dotnode[dotstyle=o](2.5,0.0){B1}%
    \dotnode[dotstyle=o](3.5,0.0){B2}%
    \dotnode[dotstyle=o](6.5,0.0){B3}%
    \dotnode[dotstyle=o](7.5,0.0){B4}%
    \dotnode[dotstyle=o](4.5,3.0){C1}%
    \dotnode[dotstyle=o](5.5,3.0){C2}%
    \dotnode[dotstyle=o](6.5,1.5){C3}%
    \dotnode[dotstyle=o](7.5,1.5){C4}%
    \ncline{->}{B1}{B2}\naput{$e_1$}%
    \ncline{->}{B3}{B4}\naput{$e_2$}%
    \ncline{->}{C1}{C2}\naput{$f_1$}%
    \ncline{->}{C3}{C4}\naput{$f_2$}%
    \ncline[linestyle=none]{-}{A1}{B1}\MyZigZagConnection[PB]{0.167}{6}%
    \ncline[linestyle=none]{-}{B2}{A2}\MyZigZagConnection[PC]{0.167}{6}%
    \ncline[linestyle=none]{-}{A2}{B3}\MyZigZagConnection[PD]{0.167}{6}%
    \ncline[linestyle=none]{-}{B4}{A3}\MyZigZagConnection[PE]{0.167}{6}%
    \myncarc[linestyle=none,arcangle=37]{-}{A1}{C1}\MyZigZagConnection[QA]{0.056}{18}%
    \myncarc[linestyle=none,arcangle=37]{-}{C2}{A3}\MyZigZagConnection[QB]{0.053}{19}%
    \myncarc[linestyle=none,arcangle=37]{-}{A2}{C3}\MyZigZagConnection[QC]{0.111}{9}%
    \myncarc[linestyle=none,arcangle=37]{-}{C4}{A3}\MyZigZagConnection[QD]{0.111}{9}%
  \end{pspicture}
\end{myfigure} 

Now, consider an instance of the game with flow over time
$\FoTInstanceName{B}=\FoTInstance$ on the network $\FoTGraph$ with the source
node $\FoTSourceNode=v_1$, the sink node $\FoTSinkNode=v_3$, and the network
input supply $\FoTSupply=\alpha_0$. The free flow transit times and the link
capacities are defined as follows\colon{}
$\FoTTransitTimeLink{e_1}=\FoTTransitTimeLink{e_2}=0$,
$\FoTTransitTimeLink{f_1}=\FoTTransitTimeLink{f_2}=T$,
$\FoTCapacityLink{e_1}=\alpha_1$, $\FoTCapacityLink{e_2}=\alpha_2$,
$\FoTCapacityLink{f_1}=\alpha_0-\alpha_1$ and
$\FoTCapacityLink{f_2}=\alpha_1$, where $T>0$ and
$0<\alpha_2<\alpha_1<\alpha_0=\FoTSupply$. Further, set the free flow transit
times of all other subdivided links to zero and the free flow transit times of
all other network links to some sufficiently big constant, say $3\cdot{}T$.
The capacities of all links except $e_1$, $e_2$, $f_1$ and $f_2$ are set to
$\alpha_0$, i.e., the amount of the network supply.

This instance restricted to the subdivided paths is equivalent to the instance
$\FoTInstanceName{A}_3$. The highest experienced transit time of a flow
particle in  every Nash equilibrium of the instance $\FoTInstanceName{A}_3$ is
arbitrarily close to $2\cdot{}T$, as shown in Lemma \ref{lma:main_lema},
but not more then $2\cdot{}T$. So, no flow particle in  any Nash equilibrium
of the instance $\FoTInstanceName{B}$ uses any non-subdivided link, as the
transit times on all non-subdivided links are much bigger. Therefore, the
highest experienced transit time of a flow particle in  every Nash equilibrium
of the instance on the entire network $\FoTGraph$ is arbitrarily close to
$2\cdot{}T$, as well.

Similarly,  all Nash equilibria of the instances $\FoTInstanceName{B}$ and
$\FoTInstanceName{A}_3$ on respective networks with the links $e_2$ removed
are equivalent, and the highest experienced transit times of flow particles in
their Nash equilibria are the same and equal to $T$. Hence, Braess's ratio of
the instance $\FoTInstanceName{B}$ is at least $2$, and so the network $G$
admits Braess's paradox.

If the network $\FoTGraph$ contains $M_3'$ or $M_3''$ as
a topological minor then the proof is analogous, with the only difference that
the free flow transit times of links subdivided from the link $g$ are set to
$0$ and their capacities are set to the network supply $\alpha_0$.\qed
\end{proof} 

\begin{conjecture}
[Necessary and sufficient condition for Braess's paradox]%
A network $\FoTGraph$ admits Braess's paradox in the model of games with
flow over time if and only if the network $\FoTGraph$ contains either the
network $M_3$, $M_3'$ or $M_3''$ as a topological
minor.
\end{conjecture} 

Another natural question to ask is, which topologies admit Braess's paradox if
we would like to use the networks in both directions. That is, if every
network could be used in the way it is defined to route the traffic from its
source to its sink, and also if we could transpose it and use it to route the
traffic in the opposite direction from the original sink to the original
source traveling along reversed links.

A network in Wardrop's model admitted Braess's paradox if and only if
its transpose admitted it. However, as we have shown, this is not the case for
the model of games with flow over time. So, for the model of games with flow
over time, we would like to characterize the class of networks which admit 
Braess's paradox either in their original or in their reverse direction.

By Theorem \ref{thm:sufficient_cond}, for every network $\FoTGraph$,
we know that either the network $\FoTGraph$ or its transpose
$\MyGraphTranspose{\FoTGraph}$ admits Braess's paradox in the model of
games with flow over time if the network $\FoTGraph$ contains either the
network $M_3$, $M_3'$ or $M_3''$ as a topological
minor or its transpose $\MyGraphTranspose{\FoTGraph}$ contains a transpose of
any of these three networks as a topological minor. This is equivalent to the
condition that the network $\FoTGraph$ contains either the network
$M_3$, $\MyGraphTranspose{M_3}$, $M_3'$ or
$M_3''$ as a topological minor, since the networks $M_3'$ and
$M_3''$ are symmetric, i.e., they are isomorphic to their transposes.

Call a network a \emph{chain of parallel paths} if it can be constructed from
a chain of parallel links by a number of link subdivisions, see
Fig.~\ref{fig:chain_of_parallel_paths} for illustration. We say that two
nodes $u$ and $v$ of a network \emph{use a chain of parallel paths} if the
union of all paths from $u$ to $v$ is  a chain of parallel paths, or there
is no path from $u$ to $v$ in the network. Further, we say that a network
\emph{uses only chains of parallel paths} if every pair of the network nodes
uses a chain of parallel paths.

\begin{myfigure} \label{fig:chain_of_parallel_paths} 
  \caption{%
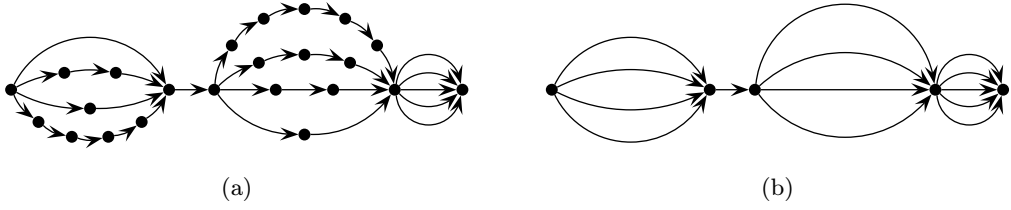
    (a) Example of a chain of parallel paths; %
    (b) and the corresponding chain of parallel links.%
  }%
  \hbox to \hsize\bgroup%
    \hfill%
    \begin{pspicture}[linewidth=0.5pt,dotscale=1.5,arrowscale=2,unit=0.6cm](0,-1.5)(6,1.3)%
      \dotnode(0,0){A1}%
      \dotnode(3.5,0){A2}%
      \dotnode(4.5,0){A3}%
      \dotnode(8.5,0){A4}%
      \dotnode(10,0){A5}%
      \psset{arrows=->}%
      \MyArcPath{25}{A1}{A2}{0.33,0.67}%
      \MyArcPath{60}{A1}{A2}{}%
      \MyArcPath{-25}{A1}{A2}{0.5}%
      \MyArcPath{-60}{A1}{A2}{0.2,0.4,0.6,0.8}%
      \ncline{A2}{A3}%
      \MyArcPath{0}{A3}{A4}{0.33,0.67}%
      \MyArcPath{40}{A3}{A4}{0.25,0.5,0.75}%
      \MyArcPath{80}{A3}{A4}{0.167,0.333,0.5,0.667,0.833}%
      \MyArcPath{-50}{A3}{A4}{0.5}%
      \MyArcPath{0}{A4}{A5}{}%
      \MyArcPath{40}{A4}{A5}{}%
      \MyArcPath{75}{A4}{A5}{}%
      \MyArcPath{-40}{A4}{A5}{}%
      \MyArcPath{-75}{A4}{A5}{}%
      \rput(5,-2.2){(a)}%
    \end{pspicture}%
    \hfill%
    \begin{pspicture}[linewidth=0.5pt,dotscale=1.5,arrowscale=2,unit=0.6cm](0,-1.5)(6,1.3)%
      \dotnode(0,0){A1}%
      \dotnode(3.5,0){A2}%
      \dotnode(4.5,0){A3}%
      \dotnode(8.5,0){A4}%
      \dotnode(10,0){A5}%
      \psset{arrows=->}%
      \MyArcPath{25}{A1}{A2}{}%
      \MyArcPath{60}{A1}{A2}{}%
      \MyArcPath{-25}{A1}{A2}{}%
      \MyArcPath{-60}{A1}{A2}{}%
      \ncline{A2}{A3}%
      \MyArcPath{0}{A3}{A4}{}%
      \MyArcPath{40}{A3}{A4}{}%
      \MyArcPath{80}{A3}{A4}{}%
      \MyArcPath{-50}{A3}{A4}{}%
      \MyArcPath{0}{A4}{A5}{}%
      \MyArcPath{40}{A4}{A5}{}%
      \MyArcPath{75}{A4}{A5}{}%
      \MyArcPath{-40}{A4}{A5}{}%
      \MyArcPath{-75}{A4}{A5}{}%
      \rput(5,-2.2){(b)}%
    \end{pspicture}%
    \hfill%
  \egroup%
\end{myfigure} 

We will show that the networks that use only chains of parallel paths are the
only networks that contain neither the network $M_3$,
$\MyGraphTranspose{M_3}$, $M_3'$ nor $M_3''$ as
a topological minor. Then we will show that no network that uses only chains
of parallel paths admits Braess's paradox in the model of games with flow
over time. This will give us a necessary and sufficient condition of existence
of Braess's paradox either in a network or in its transpose.

\begin{lemma}
\label{lma:iff_condition:lemma1} %
A network $\FoTGraph$ uses only chains of parallel paths if and only if it
does not contain any of the networks $M_3$,
$\MyGraphTranspose{M_3}$, $M_3'$ and $M_3''$ as
a topological minor.
\end{lemma} 

\begin{proof} 
If the network $\FoTGraph$ contains $M_3$ as a topological minor, then
it has a subgraph that is a subdivision of $M_3$. As $M_3$ is
not a chain of parallel paths, neither its subdivision nor the entire network
$\FoTGraph$ uses only chains of parallel paths. The same holds if the network
$\FoTGraph$ contains $\MyGraphTranspose{M_3}$, $M_3'$ or
$M_3''$ as a topological minor.

Now, let's assume the network $\FoTGraph$ does not use only chains of parallel
paths. We will prove that it contains at least one of the four networks
mentioned above as a topological minor. As $\FoTGraph$ does not use only
chains of parallel paths, it has two nodes $u$ and $v$, for which the union of
all paths from $u$ to $v$ is not a chain of parallel paths. If there are more
such pairs of nodes, take the pair with the smallest corresponding union of
paths, that is the one with minimal number of links, and denote it by $H$. The chosen
union $H$ of paths does not contain a cut vertex, that is a node that would
separate $u$ from $v$. If it contained one, say $w$, at least one of the
unions of paths from $u$ to $w$ or from $w$ to $v$ would not be a chain of
parallel paths and would be smaller than the original union of paths.

As $H$ does not contain a cut vertex, we know by Menger theorem that it
contains at lest two independent paths from $u$ to $v$. Take a maximal set of
independent paths from $u$ to $v$ in $H$ and denote it by $P$. Since $H$ is not
a chain of parallel paths, there is a path $p$ in $H$ which does not belong to
$P$, but which intersects at least one path from $P$ in a node different to
$u$ and $v$. Let $v_0$, $e_1$, $v_1$, $e_2$, $\dots$, $e_k$, $v_k$ denote the
nodes and links of $p$ in order from $u$ to $v$, respectively. Naturally
$v_0=u$ and $v_k=v$. Now we have two cases\colon{}

If the link $e_1$ does not belong to any $P$ path, then take the first node
(except $u$) of $p$ that belongs to some $P$ path. As $p$ intersects with some
$P$ path in a node different to $u$ and $v$, the graph $H$ contains
$\MyGraphTranspose{M_3}$ as a topological minor.

If $e_1$ belongs to some $P$ path, then take the first link $e_i$ of $p$ that
does not lie on this $P$ path. Such link exists, since $p\not\in{}P$, and it does
not belong to any $P$ path, since $P$ paths are independent. Let $v_j$ denote the
first node of $p$ after $e_i$ which belongs to some $P$ path. If $v_j=v$ then
$H$ contains $M_3$ as a topological minor. Otherwise, the path $p$
intersects some $P$ path again. If it is the same $P$ path as the first $p$
link belongs to, i.e., if $v_j$ and $e_1$ belong to the same $P$ path, then
$H$ contains $M_3'$ (see Fig.~\ref{fig:G_3_and_company}(c)) as
a topological minor. If it is some other $P$ path, then $H$ contains
$M_3''$ (see Fig.~\ref{fig:G_3_and_company}(d)) as
a topological minor.\qed
\end{proof} 

\begin{lemma}
\label{lma:iff_condition:lemma2} %
If a network $\FoTGraph$ uses only chains of parallel paths, then it does not
admit Braess's paradox in the model of games with flow over time.
\end{lemma} 

\begin{proof} 
Foremost, take a chain of parallel links $H$ with a source and sink nodes
$\FoTSourceNode$ and $\FoTSinkNode$, respectively, such that every $H$ link
belongs to some $\FoTST$-path. We will show that no instance
$\FoTInstanceName{B}=\FoTInstance[G=H]$ on the network $H$ admits Braess's
paradox. We have two cases\colon{}

If $H$ consists of only two nodes $\FoTSourceNode$ and $\FoTSinkNode$ and $m$
parallel links $e_1$, $e_2$, $\dots$, $e_m$ from $\FoTSourceNode$ to
$\FoTSinkNode$, then, without loss of generality, assume that
$\FoTTransitTimeLink{e_1}\leq\FoTTransitTimeLink{e_2}\leq\dots\leq\FoTTransitTimeLink{e_m}$,
and take the smallest integer $k$ such that the network supply
$\FoTSupply\leq\FoTCapacityLink{e_1}+\FoTCapacityLink{e_2}+\dots+\FoTCapacityLink{e_k}$.
It is easy to see that the flow particles in  every Nash equilibrium of this
instance begin to use the links $e_1$, $e_2$, $e_3$, and so, consecutively in
order by their free flow transit times, until they will eventually use all
links up to $e_k$ with their total capacity sufficient for the network supply.
Therefore, the inflow on the sink node never decreases and eventually
stabilizes on $\FoTSupply$. Similarly, the transit time of flow particles does
not decrease and eventually stabilizes at its maximum equal to
$\FoTTransitTimeLink{e_k}$.

If we remove a set of links from $H$, such that the new total network capacity
is still at least the network supply, the index $k$ of the most expensive used
link may not decrease, and so the maximum experienced transit time of a flow
particle in  any Nash equilibrium may not decrease as well. Therefore, the
instance $\FoTInstanceName{B}$ does not admit Braess's paradox. Note, that the
same holds even if the network supply is not just a constant $\FoTSupply$, but
also if it is a nondecreasing function that eventually stabilizes on
$\FoTSupply$. If we removed the links from $H$, such that the new total
network capacity would be strictly smaller than the network supply, then the
maximum experienced transit time of a flow particle in the new network would
be unbounded. Hence, it would not decrease.

If $H$ contains $n$ nodes, $n\geq{}3$, then it is a chain of $n-1$ sets of
parallel links. Denote its nodes by $\FoTSourceNode=v_1$, $v_2$, $\dots$,
$v_n=\FoTSinkNode$ in order of their distance from $\FoTSourceNode$. We can
decompose $H$ into $n-1$ subgraphs such that the subgraph $H_k$ contains the
nodes $v_k$ and $v_{k+1}$ and the links connecting them, where $1\leq{}k<n$.
Every flow particle traverses the subgraphs in the fixed order. Moreover, in
 every Nash equilibrium, any two flow particles that enter the subgraph $H_k$ at
the same moment leave it at the same moment as well, as they both choose only
the currently shortest paths from $v_k$ to $v_{k+1}$. So, every pair of flow
particles that enters the network $H$ together, enters every subgraph $H_k$
together. Hence, in  every Nash equilibrium of $\FoTInstanceName{B}$, the
behaviour of all flow particles in the subgraph $H_k$ is equivalent to their
behaviour in  some Nash equilibrium of  the instance
$\FoTInstanceName{B}_k=\FoTInstance[G=H_k,s=v_k,t=v_{k+1},d=\FoTSupply(\theta)]$
restricted to the subgraph $H_k$ with the network supply function
$\FoTSupply(\theta)$ equal to the outflow on the node $v_k$.

From the case for chains of parallel links with only two nodes, we know that
the transit times of flow particles in  every Nash equilibrium of
$\FoTInstanceName{B}_k$ eventually stabilizes on its maximum value. So, the
maximum experienced transit time of a flow particle in  every Nash equilibrium of
$\FoTInstanceName{B}$ is the sum of maximum experienced transit times of flow
particles in Nash equilibria of all $\FoTInstanceName{B}_k$. As no removal of
any $H_k$ links may cause any decrease of the maximum experienced transit time
in  any Nash equilibrium of $\FoTInstanceName{B}_k$, it may not decrease the
maximum experienced transit time in  any Nash equilibrium of
$\FoTInstanceName{B}$ as well. Therefore the instance $\FoTInstanceName{B}$
does not admit Braess's paradox.

Now, take a network $G$ that uses only chains of parallel paths.
We will show that no instance of a game with flow over time on the
network $\FoTGraph$ admits Braess's paradox. So, take an instance
$\FoTInstanceName{A}=\FoTInstance$ of such a game on the network $\FoTGraph$.
As $\FoTGraph$ uses only chains of parallel paths, the union of all paths from
$\FoTSourceNode$ to $\FoTSinkNode$ is a chain of parallel paths, denote it by
$H$. As no flow particle of the instance $\FoTInstanceName{A}$ may use any
non-$H$ link, the instance $\FoTInstanceName{A}$ admits the paradox if and
only if its restriction $\FoTInstanceName{A'}=\FoTInstance[G=H]$ to the
subgraph $H$ admits it as well.

Note that if a graph $X'$ is created from a graph $X$ by a link smoothing (an
inverse operation to the link subdivision), in which two consecutive $X$ links
$e_1$ and $e_2$ are smoothed out into one $X'$ link $e$, such that
$\FoTTransitTimeLink{e}'=\FoTTransitTimeLink{e_1}+\FoTTransitTimeLink{e_2}$
and
$\FoTCapacityLink{e}'=\min\{\FoTCapacityLink{e_1},\FoTCapacityLink{e_2}\}$,
assuming that neither the network source nor the network sink node was
smoothed out, then the instances
$\FoTInstance[G=X',tau=\FoTTransitTime',c=\FoTCapacity']$ and
$\FoTInstance[G=X]$ are equivalent in a sense that all flow particles behave
the same way in both instances. Therefore, both instances admit the paradox if
and only if the other one admits it as well.

So, take an instance
$\FoTInstanceName{B}=\FoTInstance[G=H',tau=\FoTTransitTime',c=\FoTCapacity']$
on a network $H'$ created from $H$ by a maximal number of link smoothings
possible, with the free flow transit times and link capacities adjusted
appropriately. As $H$ is a union of $\FoTST$-paths in $\FoTGraph$, no link
enters $\FoTSourceNode$ nor leaves $\FoTSinkNode$, and thus it is not possible
to smooth $\FoTSourceNode$ nor $\FoTSinkNode$ out. Therefore, both nodes
$\FoTSourceNode$ and $\FoTSinkNode$ remain in $H'$. Moreover, $H'$ is a chain
of parallel links, since $H$ is a chain of parallel paths, however, we know
that no instance on a chain of parallel links admits Braess's paradox.
Therefore, according the previous paragraph, neither the instance
$\FoTInstanceName{B}$ nor the instance $\FoTInstanceName{A}$ admits it.\qed
\end{proof} 

Thus, from the previous lemmas and the fact that a network admits Braess's
paradox in the model of games with flow over time if it contains either $M_3$,
$M_3'$ or $M_3''$ as a topological minor, we get the following theorem\colon{}

\begin{theorem}
[Necessary and sufficient condition for Braess's paradox both-ways]%
For any network $\FoTGraph$, the following statements are equivalent\colon{}
\begin{enumerate}
\item[(i)]%
Either the network $\FoTGraph$ or its transpose $\MyGraphTranspose{\FoTGraph}$
admits Braess's paradox in the model of games with flow over time.
\item[(ii)]%
The network $\FoTGraph$ contains either $M_3$,
$\MyGraphTranspose{M_3}$, $M_3'$ or $M_3''$ as
a topological minor.
\item[(iii)]%
The network $\FoTGraph$ does not use only chains of parallel paths.
\end{enumerate}
\end{theorem} 


\section{Conclusion} 

We have proved several new properties of Braess's paradox for congestion games
with flow over time. However, a number of questions have been left open.

We showed that there are networks which do not admit Braess's paradox in games
with static flows, but which admit it in the model with flow over time. We
showed that these networks admit a  much more severe Braess's ratio
for this model. In particular, we showed that Braess's ratio of the class of
all instances of games with flow over time on networks with $n$ nodes is at
least $n-1$. What is the upper bound on Braess's ratio for this model?

Then, we illustrated that Braess's paradox is not symmetric for flows over
time, although it is symmetric for the case of static flows. We showed
that there are network topologies which exhibit Braess's paradox, but for
which the transpose does not. Is this asymmetry of Braess's paradox inherent
for flows over time? What are the properties of Braess's paradox for different
models of games with flows over time?

Finally, we conjectured a necessary and sufficient condition of existence of
Braess's paradox in a network, and proved the condition of existence of the
paradox either in the network or in its transpose. Is this conjecture valid in general?


\bibliography{bibliography}%
\bibliographystyle{abbrv}%

\end{document}